\documentclass[11pt]{article}
\usepackage{amsmath}
\usepackage{amsthm}
\usepackage{amsfonts}
\usepackage{bbm}
\usepackage{graphicx}
\usepackage{sectsty}
\RequirePackage{setspace}
\usepackage{appendix}
\usepackage{float}
\usepackage{url}
\usepackage{caption}
\usepackage{lineno}
\usepackage[round]{natbib}

\newcommand{\captionfonts}{\footnotesize}
\makeatletter % Allow the use of @ in command names
\long\def\@makecaption#1#2{%
\vskip\abovecaptionskip
\sbox\@tempboxa{{\captionfonts #1: #2}}%
\ifdim \wd\@tempboxa >\hsize
{\captionfonts #1: #2\par}
\else
\hbox to\hsize{\hfil\box\@tempboxa\hfil}%
\fi
\vskip\belowcaptionskip}
\makeatother 

\begin{document}
\title{{\bf Quantum entanglement in physical and cognitive systems: a conceptual analysis and a general representation}}
\author{Diederik Aerts$^*$,\, Jonito Aerts Argu\"elles$^*$,\, Lester Beltran$^*$,\, Suzette Geriente\footnote{Center Leo Apostel for Interdisciplinary Studies, Brussels Free University, Krijgskundestraat 33, 1160 Brussels (Belgium). Email addresses: \emph{diraerts@vub.ac.be,jonitoarguelles@gmail.com,lbeltran@vub.ac.be,sgeriente@vub.ac.be,msassoli@vub.ac.be}} \\ Massimiliano Sassoli de Bianchi$^*$\footnote{Laboratorio di Autoricerca di Base, via Cadepiano 18, 6917 Barbengo (Switzerland). Email address:\emph{autoricerca@gmail.com}}\,\, Sandro Sozzo\footnote{School of Business and Research Centre IQSCS, University Road, LE1 7RH Leicester (United Kingdom). Email address: \emph{ss831@le.ac.uk}}\, and Tomas Veloz\footnote{Universidad Andres Bello, Departamento Ciencias Biol{\' o}gicas, Facultad Ciencias de la Vida, 8370146 Santiago (Chile), Instituto de Filosof{\' i}a y Ciencias de la Complejidad IFICC, Los Alerces 3024, \~Nu\~noa, Santiago (Chile). Email address: \emph{tveloz@gmail.com}}\ $^*$}

\date{}

\maketitle

\begin{abstract}
\noindent
We provide a general description of the phenomenon of entanglement in bipartite systems, as it manifests in micro and macro physical systems, as well as in human cognitive processes. We do so by observing that when genuine coincidence measurements are considered, the violation of the `marginal laws', in addition to the Bell-CHSH inequality, is also to be expected. The situation can be described in the quantum formalism by considering the presence of entanglement not only at the level of the states, but also at the level of the measurements. However, at the ``local'' level of a specific joint measurement, a description where  entanglement is only  incorporated in the state remains always possible, by adopting a fine-tuned tensor product representation. But contextual tensor product representations should only be considered when there are good reasons to describe the outcome-states as (non-entangled) product states. This will not in general be true, hence, the entangement resource will have to generally be allocated both in the states and in the measurements. In view of the numerous violations of the marginal laws observed in physics' laboratories, it remains unclear to date if entanglement in micro-physical systems is to be understood only as an `entanglement of the states', or also as an `entanglement of the measurements'. But even if measurements would also be entangled, the corresponding violation of the marginal laws (no-signaling conditions) would not for this imply that a superluminal communication would be possible. 
\end{abstract}
\medskip
{\bf Keywords:} Human cognition; quantum structures; quantum measurements; Bell inequalities; entanglement; superposition; marginal laws; no-signaling conditions; marginal selectivity. 
%%SS I would avoid explicitly mentioning ''violation of the non-signaling conditions" at the beginning of the paper. Can we use "marginal laws" instead? 
%%SS In the keynotes, I would take out "marginal selectivity". Also, I would write "Bell inequalities", taking out both "Bell inequality" and "CHSH inequality".

\section{Introduction\label{intro}}

The term `entanglement' was firstly officially introduced by Schr\"{o}dinger, in the thirties of the last century. He described it as a situation of two systems whose states are initially known, which following a temporary interaction enter into a state where one has a complete knowledge of the state of the bipartite system formed by their combination, but not anymore (apparently at least) a complete knwowledge of their individual states, and this even though they may become widely separated in space and therefore, one would expect, in a condition such that they should possess well-defined individual properties. Schr\"{o}dinger famoulsy asserted that he did not consider entanglement as ``\emph{one} but rather \emph{the} characteristic trait of quantum mechanics, the one that enforces its entire departure from classical lines of thought'' \citep{Schrodinger1935}.

Thirty years later, John Bell derived the famous inequalities that today bear his name, which are able to test the presence of entanglement in bipartite systems \citep{Bell1964}. When he did so, there was still an ongoing and widespread debate in the physics' community regarding the validity of quantum mechanics as a fundamental theory of our physical reality. Bell himself did not believe that in actual experiments, like those that would be realized by Aspect's and other experimental groups in the following decades, his inequalities would be violated \citep{aspect1982a,aspect1983,aspect1982b,tittel1998,weihs1998,vienna2013,urbana2013,hensen-etal2016}. But nowadays the predictions of quantum theory are no longer put into question, not only as regards entanglement, which has been shown to be preservable over distances of more than a thousand of kilometers \citep{Yin2017}, but also with respect to many other effects predicted by the theory, like the delocalization of large organic molecules \citep{Gerlich2011}, just to cite one. On the other hand, the debate about the profound meaning of the theory never stopped, and in fact has constantly renewed and expanded over the years, so much so that one can envisage this will produce in the end a Copernican-like revolution in the way we understand the nature of our physical reality \citep{Deutsch1998,Stapp2011,k2013,Fuchs2017,Aertsetal2018}. Such debate, however, has not remained confined to physicists or philosophers of science, but also reached new fields of investigation, in particular that of psychology, due to the development of that research domain called `quantum cognition', which saw its beginnings in the nineties of the last century  \citep{aertsaerts1995,aertsbroekaertsmets1999,khrennikov1999,gaboraaerts2002,altmanspacher2002,aertsczachor2004,AertsGabora2005a,AertsGabora2005b} and borrowed ideas from quantum physics to develop new promising models for a variety of cognitive phenomena, also providing in return interesting insights as regards our understanding of physical systems \citep{Khrennikov2010,BusemeyerBruza2012,HavenKhrennikov2013,Wendt2015}.

That said, it is worth observing that since the days of Schr\"{o}dinger, one of the main elements of dissatisfaction was the presence of an irreducible (irremovable) probability in quantum theory. This famously led Einstein to say, in a letter that he wrote to Max Born in 1926, that God ``is not playing at dice'' \citep{Born1971}. Following the research that was carried out during the last four decades by our group, first in Geneva and then in Brussels, we can say in retrospect that we do agree with him, as we are today confident in asserting that quantum mechanics is not at odds with determinism, if the latter is understood at the global level \citep{Aerts1999b,AertsSassolideBianchi2014,AertsSassoli2016}. God does not play dice, indeed, because s/he does not have to, as the irreducible quantum probabilities come from the fact that, as we will explain in the article, as humans we are forced to play dice when we perform quantum measurements, being the latter much like (weighted) symmetry breaking processes (the actual breaking the symmetry of the potential), integrating in their very protocol the presence of fluctuations that cannot be eliminated without at the same time altering the very nature of what is being measured \citep{SassolideBianchi2015}. 

The view that quantum entities might not always have well-defined values for certain observables, and this not because we would be ignorant about these values, but because, literally, there would be no actual values, was considered to be problematic by many, as against the view of realism, although this is truly a matter of concern only for those adhering to the (we think false) prejudice that our physical reality should be fully contained in space (or spacetime). However, considering all that
we learned from quantum physics and relativity, this is very likely to be a wrong (or incomplete) view, space being instead only a very particular theater staging a small portion of our multidimensional reality, typically that populated by the classic macroscopic entities \citep{Aerts1999b,AertsSassolideBianchi2014,AertsSassolideBianchi2017c}.

The presumed antirealism of quantum theory has brought people to investigate whether it would be possible to substitute quantum theory by so called hidden-variable theories, aiming at explaining the quantum probabilities as resulting from our lack of knowledge of an underlying (pre-empirical, pre-spatial, non-spatial) deterministic reality. Bell's work, in the sixties of last century \citep{Bell1964,Bell1966,Bell1987}, fits into this context of searching for a hidden-variable theory \citep{genovese2005}, and was deeply inspired by a situation that was described in 1935 by Einstein, Podolsky and Rosen (EPR), still called today the `EPR paradox' \citep{epr1935}, although there are no more paradoxes, the situation having been clarified not only from the experimental point of view \citep{aspect1982a,aspect1983,aspect1982b,tittel1998,weihs1998,vienna2013,urbana2013,hensen-etal2016}, but also from the logical one \citep{a1984,Sassoli2019}. 

The situation put forward in the EPR article was later reformulated by David Bohm, using the clearer example of two spins in an entangled spin state \citep{Bohm1951}, which is today considered to be the archetypical quantum entanglement situation, and we will also consider such example in our discussion in this article. Regarding how the Bohm model situation is described by many working physicists, we observe that there is still a disturbing ``schizophrenia'' about how the two entangled spin entities in a Bohm-EPR set up are viewed. On the one hand, there is agreement in acknowledging that two spins, even though separated in spatial terms, nevertheless form a single and whole system. On the other hand, there is difficulty in accepting the consequence of such statement, implying that our spatial theater, as we said already, can only be viewed as the tip of a much vaster non-spatial reality, which cannot be fully represented in space, hence, cannot be understood only in terms of spatio-temporal phenomena akin to localized particles or extended waves and fields. 

In another letter to Max Born, Einstein wrote in 1947 \citep{Born1971}: ``I admit, of course, that there is a considerable amount of validity in the statistical approach which you were the first to recognise clearly as necessary given the framework of the existing formalism. I cannot seriously believe in it because the theory cannot be reconciled with the idea that physics should represent a reality in time and space, free from spooky actions at a distance. [$\cdots$] I am quite convinced that someone will eventually come up with a theory whose objects, connected by laws, are not probabilities but considered facts, as used to be taken for granted until quite recently.'' In a later commentary, Born wrote that the decisive sentence in Einstein's letter \citep{Born1971}: ``[$\cdots$] is the one where he says: `that physics should represent a reality in time and space, free from spooky actions at a distance'. I too had considered this postulate to be one which could claim absolute validity. But the realities of physical experience had taught me that this postulate is not an \emph{a priori} principle but a time-dependent rule which must be, and can be, replaced by a more general one.''

This more general rule, mentioned by Born, asks us to abandon the `space contains reality hypothesis' and to accept what follows from the quantum formalism and its numerous successful tests in the laboratories, i.e., that \citep{Aerts1999b}: ``reality is not contained within space. Space is a momentaneous crystallization of a theatre for reality where the motions and interactions of the macroscopic material and energetic entities take place. But other entities -- like quantum entities for example -- `take place' outside space, or -- and this would be another way of saying the same thing -- within a space that is not the three-dimensional Euclidean space.''

Ironically enough, Einstein's reality criterion \citep{epr1935} provides one of the strong arguments that intimate us to accept the non-spatiality of the quantum micro-entities. Indeed, Heisenberg's uncertainty principle prevents us to simultaneously define both the position and momentum of a quantum entity like an electron. Therefore, one cannot determine, not even in principle, how the position and momentum of the entity will vary in time, and consequently one cannot predict with certainty, not even in principle, its future locations. Following Einstein's reality criterion, we have then to conclude that the entity in question doesn't possess the property of being somewhere in space, hence it would be a non-spatial entity, which does not mean, however, that it would be an unreal entity \citep{Sassoli2011}.

Many additional arguments can be brought forward in support of the thesis that quantum entities should be considered to be non-spatial, like those following from a study of their temporal behaviors, by means of the notion of sojourn time \citep{Sassoli2012}, or from an analysis of spins greater than one-half, which cannot be associated with any specific spatial direction \citep{AertsSassolideBianchi2017c}, and of course, there are also the many no-go theorems, in particular those of \citet{Kochen1967}, which if taken seriously tell us just that: that quantum entities cannot be depicted as the factual objects connected by laws of Einstein's desiderata, being instead more like entities having an unexpected ``conceptual nature,'' being able to manifest in states having a varying degree of abstractness or concreteness, the more concrete ones being those we usually describe as the classical spatio-temporal objects of our ordinary experience \citep{Aertsetal2018}. 

It is certainly not the purpose of the present paper to enter into a comprehensive discussion of the non-spatial and conceptual behavior of quantum entities (see also, in that respect, the perspective offered by Kastner's possibilistic transactional interpretation of quantum mechanics \citep{k2013}). We only want here to emphasize that it would be wrong to consider that a physical entity, to be real, has to exist in space. If we let go such ``classical prejudice,'' then, when studying the phenomenon of entanglement, one is not forced any more, as Einstein considered to be, to understand two entities that are separated in space as two entities that would be necessarily disconnected to one another. If their state is non-spatial, then the nature of their possible connection will simply be non-spatial as well, i.e., non-manifest as a connection through space. If this is so, then there is no need to speak of ``spooky actions at a distance,'' as there would be no phantom-like action of one entity over the other, during a Bell-test experiment. More simply, Alice and Bob,\footnote{Alice and Bob are the two archetypical fictional characters used to described, in physics and cryptology, the joint experimental actions and data collections resulting from their sharing of a same bipartite system, typically in an entangled state.} with their instruments, localized in different regions of space, would both act on a same entity, which forms a whole, and not on two independent entities. 

In other words, it would be wrong to conceive a bipartite entity in an entangled state as two fully separated entities, just because they can respond to different instruments placed at distant locations, and that in between these locations the probability of a detection tends to zero, once the entangled entities have been emitted by the source and have propagated away. If outcomes can be actualized in a coincident and correlated way, in distant detectors, this is because the two apparent spatially divided fragments are the tip of an undivided non-spatial entity, having some well-defined degree of availability in interacting with different spatial measuring instruments, and by doing so acquiring spatial properties (for instance by leaving a trace in a detection screen). If this is correct, then there is no reason to speak
in terms of an action, or influence, of Alice's measurement on Bob's measurement, and vice versa, as they would both operate on a same undivided entity, at the same time. There is no influence of one measurement on the other, only a single measurement jointly performed by Alice and Bob, on a whole entity. The latter can then remain whole, or possibly disentangle, following their joint action, depending on the entity's nature and on the experimental operations. When joint measurements on entangled entities are understood in this way, there is also no reasons to require that Alice's statistics of outcomes would be independent of the choice of actions operated by Bob, i.e., that the system would necessarily obey the so-called `no-signaling conditions' (also called `marginal laws', or `marginal selectivity'). 

These no-signaling conditions are implicitly assumed to be valid in the standard quantum formalism, when joint measurements are represented by 
tensor product observables, defined with respect to a same given tensor product decomposition of the Hilbert space. However, it remains today unclear if this is the correct way to model certain experimental situations, considering that significant violations of the no-signaling conditions have been evidenced in the physics' laboratories 
\citep{AdenierKhrennikov2007,DeRaedt2012,DeRaedt2013,AdenierKhrennikov2016,Bednorz2017,Kupczynski2017}. These violations are totally unexpected if the adopted view is that in which Alice and Bob perform independent (although coincident in time) measurements, instead of a bigger unique joint measurement. In other words, if we think of Alice's and Bob's measurements as two distinct interrogative contexts, asking different and independent questions, then there is indeed no reasons to expect that the statistics of answers collected by Alice could be influenced by the questions that are asked by Bob. 

The usual understanding of quantum entanglement in physics is indeed that of a situation where we have a bipartite entity emitted by a source, like the two spin one-half fermions of Bohm's archetypical model, such that when they are flying apart, they are assumed to become fully independent. If this would be true, it would be natural to think that the choice of measurement performed by Alice on its sub-system could not be influenced by the choice of measurement performed by Bob, and vice versa. If the two fermionic entities are fully separated (spatially and experimentally) this is indeed something to be expected, which then translates in the constraints of the no-signaling conditions. The same reasoning would however lead one to also expect that no correlations of the kind able to violate Bell's inequality should be observed between Alice's and Bob's answers, which is why Einstein described, and many physicists nowadays still describe, the situation as a ``spooky action at a distance.'' The action is considered to be ``spooky'' because no force field seems to be involved in it, but would just happen as a consequence of the linear structure of the quantum state space. 

But it seems that there is a limit in the ``spookiness'' that many physicists are ready to digest: it can be spooky enough to violate Bell's inequalities, but not so spooky to also violate the marginal laws. Why? Because these laws are the conditions that guarantee that no faster than light communications can arise between Alice and Bob. Still, the very idea of Alice sending a signal to Bob, or vice versa, relies on Alice and Bob selectively acting only on their sub-entities, which in turn presupposes some sort of separateness of these two sub-entities, which is only reasonable to assume if they would be genuine spatial entities, as then their separation in space would be sufficient to also produce a ``separation in substance.'' As we emphasized, this is however an untenable assumption, hence not only the violation of Bell's inequalities is to be expected, but also a violation of the no-signaling conditions should be expected, unless the system possesses some remarkable symmetries (which can indeed be the case for the quantum micro-systems, in case the observed violations of the no-signaling conditions would only be the result of experimental errors). This does not imply, however, that there would be the possibility to exploit these violations in order to produce a superluminal communication between Alice and Bob. As we will explain in this article, we believe that a subtle logical error has been made in the standard analysis of the no-signaling conditions. It is usually considered that since the produced correlations in a typical Bell-test situation are spacelike separated, if they could be used to send signals, then such signals would travel faster than light, hence they would violate relativity theory. The error in question consists in not carefully distinguishing between the `origin of the correlations' and the `mechanism of signaling by means of such correlations'. Indeed, even if correlations are associated with spacelike separated events, the mechanism of using them for signaling will not necessarily lead to a faster than light propagation.

What above described leads us to consider another domain of investigation, that of human cognition, where it was also observed that experimental situations can be created where Bell's inequalities are violated \citep{bkmm2008,bknm2009,as2011,as2014,bkrs2015,gs2017,aabgssv2018a,aabgssv2018b,Beltran2018,abgs2019}. This is so because there can be more or less strong (non-spatial) connections between different conceptual entities, depending on how much meaning they share. In other words, meaning connections in the conceptual realm are the equivalent of the (non-spatial) coherence-connection shared by entangled micro-physical entities, and can be exploited to create correlations in well-designed psychological experiments. Generally, these experiments will also violate the marginal laws, hence, in their modeling one cannot simply use the standard quantum representation, with a single tensor product representation for all considered joint measurements. 

This caused some authors to doubt that a genuine form of entanglement is at play in cognitive systems \citep{Dzhafarov2013,Aerts2014,aabgssv2018b}. For instance, in the abstract of \citet{CervantesDzhafarov2018}, the authors write: ``All previous attempts to find contextuality in a psychological experiment were unsuccessful because of the gross violations of marginal selectivity in behavioral data, making the traditional mathematical tests developed in quantum mechanics inapplicable.'' In our opinion, statements of these kind rest on the aprioristic view that entanglement should be caused by some sort of (``spooky'') information flow from Alice to Bob, or vice versa, rather than by a process where Alice and Bob are able to jointly co-create information/meaning, by acting at once on a same whole entity, the latter being of course a process that is expected to also generally violate the marginal laws. So, we could say that, in a sense, the same kind of misunderstanding
seems to be at play in the analysis of the entanglement phenomenon both in the physical and psychological laboratories, when some physicists and psychologists try to figure out what could be able to generate the observed correlations. 

It is the purpose of the present paper to bring some clarity to all this, presenting the phenomenon of entanglement and its quantum modeling, both in physical and cognitive situations, under the light of a unified and general perspective, resulting from the work that our group has carried out in the last decades, aimed at understanding the foundations of physical theories and of our human cognitive processes, always bringing particular attention to the aspects that unite these two domains. To do so, in Sec.~\ref{bipartite}, we start by describing three paradigmatic examples of bipartite systems violating the CHSH version of Bell's inequality, which we will use to exemplify our analysis throughout the paper. The first one (Sec.~\ref{Bohm}) is Bohm's spin model of two spin-${1\over 2}$ entities in a singlet state. The second one (Sec.~\ref{Polyelastic}) is a variation of Aerts' ``vessels of water'' macroscopic model, which involves an elastic entity formed by multiple bands. The third example (Sec.~\ref{psymodel}) is a `psychological model', exploiting the meaning connection characterizing a given conceptual combination. 

In Sec.~\ref{marginal}, we then provide a probabilistic definition of sub-measurements, and make explicit the marginal laws (or no-signaling conditions), showing that they are obeyed in Bohm's spin model example, because of the assumed product (non-entangled) structure of the measurements within the customary tensor product description of the situation, but violated in the elastic and the psychological models. In Sec.~\ref{compatibility}, we also define the notions of compatibility and separability, showing that when all joint measurements are formed by separate sub-measurements, both the CHSH inequality and the marginal laws are necessarily obeyed. In Sec.~\ref{correlations-section}, we continue our analysis by introducing and explaining the important distinction between correlations of the first and second kind, showing that only the latter, which operate at the (non-local) level of the whole entangled entity, are able to violate the CHSH inequality, without for this implying a superluminal influence traveling from Alice to Bob, or vice versa, i.e., with no ``spooky action at a distance'' and no violations of relativistic principles, even if the marginal laws are also violated.

In Sec.~\ref{Hilbert model}, we show that the standard quantum formalism allows one to model the joint measurements even in situations where the marginal laws are disobeyed, emphasizing that a tensor product structure is always relative to the choice of a specific isomorphism and that when the marginal laws are violated a single isomorphism is insufficient to introduce a tensor product structure for all joint measurements. Hence, some will necessary appear to be entangled measurements. In Sec.~\ref{Elastic modeling}, a specific modeling example is provided, and in Sec.~\ref{product-entanglement} the issue of how and when to introduce isomorphisms in order to tensorialize specific joint measurements is discussed, emphasizing that it should be limited to those situations where there is evidence that the outcome-states are product states, i.e., describe a disentanglement of the system as produced by the measurements. 

In Sec.~\ref{more-on-submeasurements}, we lean on the problem of the definition of sub-measurements when also the change of state induced by them, as described by the quantum projection postulate, is taken into due consideration, emphasizing that in quantum mechanics sub-measurements do not arise as a simple procedure of identification of certain outcomes. Finally, in Sec.~\ref{more-on-submeasurements}, we offer a number of concluding remarks, trying to bring attention to some important points that emerged from our analysis.

\section{Bipartite systems and joint measurements}
\label{bipartite}

Our discussion in this article will be limited to a particular class of systems called `bipartite systems', or `bipartite entities'. As the name indicates, these are systems (or entities, we will use these two terms interchangeably in the article) of a composite nature, i.e., in which two parts, called sub-systems, or sub-entities, can be identified at some level. When considering bipartite entities, one can adopt a ``deconstructivist viewpoint,'' where the starting point is a single whole entity and one considers to which extent such entity can be understood as a system formed by two parts, or a complementary ``constructivist viewpoint,'' where one starts from two clearly distinguishable entities that are brought together in some way, and one considers to which extent they can be understood as the sub-entities forming a bigger emerging composite system. 

Independently of the viewpoint considered, what is important for our analysis is that the system possesses some properties that are characteristics of a bipartite structure. For example, two electrons, even when in a singlet (entangled) state, form a bipartite system because even though we cannot attach individual vector-states to each one of the electrons (one can nevertheless attach individual density operator-states \citep{AertsSassoli2016}), there are properties that remain always actual and characterize the bipartiteness of the system, like the fact that we are in the presence of two electronic masses and two electric charges, that the system can produce two distinct detection events on spatially separated screens, instead of a single one, etc. In the present discussion, we will limit our analysis to three different kinds of bipartite systems, which will help us to illustrate the different aspects of our approach to quantum entanglement and its modeling. The first system is David Bohm's archetypical example of two half-spins in an entangled state. The second system is a macroscopic elastic structure presenting two distinguishable ends, on which it is possible to act simultaneously, and the third one is a combination of two abstract concepts, subjected to participants in a psychological experiment.

\subsection{Bohm's spin model}
\label{Bohm}

In the historical discussion of their paradoxical situation, EPR considered the position and momentum observables. As we mentioned in the Introduction, David Bohm subsequently proposed a simpler situation, which expresses the situation equally well: that of two spin-${1\over 2}$ entities (two fermions) in a rotationally invariant entangled state (a so-called singlet state), the wave function of which can be written as: 
\begin{equation}
\psi({\bf r}_1,{\bf r}_2)|s\rangle={1\over \sqrt{2}}\psi({\bf r}_1)\psi({\bf r}_2)(|+\rangle\otimes|-\rangle-|-\rangle\otimes |+\rangle,
\label{singletstate}
\end{equation}
where $\psi(\bf{r}_1)$ and $\psi(\bf{r}_2)$ are the spatial components of the wave functions of the two fermions, and $|+\rangle$ and $|-\rangle$
are the ``up'' and ``down'' eigenstates of the spin operators relative to some given spatial direction. The spatial component $\psi({\bf{r}_1},{\bf{r}_2})$ is of course important in order to describe the evolution of the composite entity in relation to space. However, we will focus here only on the spinorial component, being understood that the spatial factor describes two entities emitted by a source that propagate away from each other, with some average velocity, in opposite directions, towards some distant Stern-Gerlach apparatuses and the associated detection screens. So, we will more simply consider that the state of the bipartite system is:
\begin{equation}
|s\rangle={1\over \sqrt{2}}(|+\rangle\otimes|-\rangle -|-\rangle\otimes|+\rangle.
\label{singletstate2}
\end{equation}
It is worth observing that the above spin vector-state, being rotationally invariant, cannot be associated with well-defined individual spin properties prior to the measurements (we will come back to this point later in our discussion).

\begin{figure}[!ht]
\centering
\includegraphics[scale =0.24]{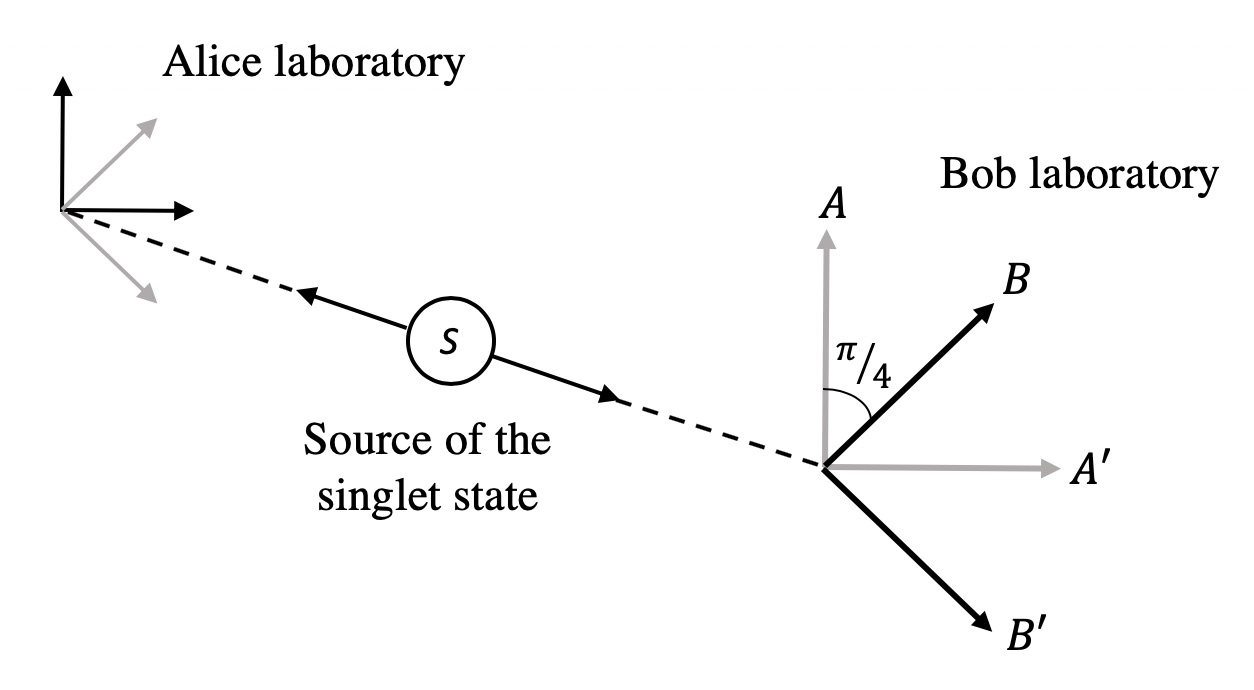}
\caption{The four orientations of the Stern-Gerlach apparatuses used by Alice and Bob in their joint measurements, which maximize the violation of the CHSH inequality.}
\label{figure1}
\end{figure}

In the typical experimental setting of a Bell-test experiment, one has four joint measurements, which we simply denote $AB$, $AB'$, $A'B$ and $A'B'$. In the joint measurement $AB$, the spin entity moving to the left (let us call it the spin measured by Alice) is subjected to a Stern-Gerlach apparatus oriented along the $A$-axis, whereas the spin entity moving to the right (let us call it the spin measured by Bob) is subjected to a Stern-Gerlach apparatus whose magnet is oriented along the $B$-axis. The same holds for the other three joint measurements, which use Stern-Gerlach apparatuses also oriented along the $A'$ and $B'$ axes. If $\alpha$ is the angle between the $A$ and $B$ axes, then according to the quantum formalism the probabilities $p(A_1,B_1)$ and $p(A_2,B_2)$ that Alice and Bob jointly obtain a spin up outcome, respectively jointly obtain a spin down outcome, are: 
\begin{eqnarray}
&p(A_1,B_1)=p(A_2,B_2)={1 \over 2}\sin^2{\alpha \over 2}.
\end{eqnarray}
On the other hand, the probabilities $p(A_1,B_2)$ and $p(A_2,B_1)$ of Alice finding a spin up and Bob a spin down, respectively Alice finding a spin down and Bob a spin up, are given by: 
\begin{eqnarray}
&p(A_1,B_2)=p(A_2,B_1)={1 \over 2}\sin^2{\pi -\alpha \over 2}={1 \over 2}\cos^2{\alpha \over 2}.
\end{eqnarray}
If an angle $\alpha={\pi\over 4}$ is considered, one then finds:
\begin{eqnarray}
&p(A_1,B_1)=p(A_2,B_2)={1\over 8}(2-\sqrt{2}),\quad p(A_1,B_2)=p(A_2,B_1)={1\over 8}(2+\sqrt{2}).
\end{eqnarray}
An optimal choice for the $A'$ and $B'$ measurements, maximizing the correlations, is to also consider an angle of ${3\pi\over 4}$ between the $A$ and $B'$ axes, and an angle of ${\pi\over 4}$ between the $B$ and $A'$ axes, with angles of ${\pi\over 2}$ between $A$ and $A'$ and between $B$ and $B'$; see Fig.~\ref{figure1}. 
This gives the additional probabilities: 
\begin{eqnarray}
&p(A_1,B'_2)=p(A_2,B'_1)=p(A'_1,B_1)=p(A'_2,B_2)=p(A'_1,B'_1)=p(A'_2,B'_2)={1\over 8}(2-\sqrt{2}),\nonumber\\
&p(A_1,B'_1) =p(A_2,B'_2)= p(A'_1,B_2)=p(A'_2,B_1)=p(A'_1,B'_2)=p(A'_2,B'_1)={1\over 8}(2+\sqrt{2}).
\end{eqnarray}

By attributing the value $+1$ to the situation where both spins are either up or down, and the value $-1$ to the situation where one spin is up and the other spin is down, we can calculate the following `expectation values' (also called `correlation functions') of the four joint measurements: 
\begin{eqnarray}
&E(A,B)=p(A_1,B_1)-p(A_1,B_2)-p(A_2,B_1)+p(A_2,B_2)=-{1\over \sqrt{2}}\nonumber\\
&E(A,B')=p(A_1,B'_1)-p(A_1,B'_2)-p(A_2,B'_1)+p(A_2,B'_2)={1\over \sqrt{2}}\nonumber\\
&E(A',B)=p(A'_1,B_1)-p(A'_1,B_2)-p(A'_2,B_1)+p(A'_2,B_2)=-{1\over \sqrt{2}}\nonumber\\
&E(A',B')=p(A'_1,B'_1)-p(A'_1,B'_2)-p(A'_2,B'_1)+p(A'_2,B'_2)=-{1\over \sqrt{2}}.
\label{correlations}
\end{eqnarray}
The Clauser Horne Shimony Holt (CHSH) version of Bell’s inequality then says that the quantity \citep{Clauser1969}:
\begin{equation}
{\rm CHSH}\equiv E(A,B)-E(A,B')+E(A',B)+E(A',B'),
\label{CHSH-quantity}
\end{equation}
or similar expressions obtained by interchanging the roles of $A$ and $A'$ and/or the roles of $B$ and $B'$, is bounded by: 
\begin{equation}
-2\leq {\rm CHSH}\leq 2.
\label{CHSH-inequality}
\end{equation}
Since (\ref{correlations}) gives ${\rm CHSH}=-{4\over \sqrt{2}}=-2\sqrt{2}<-2$, the CHSH inequality (\ref{CHSH-inequality}) is clearly violated by Bohm's spin model. The violation corresponds here to the value known as Tsirelson's bound, which is a maximal value for the quantum correlations, for as long as the no-signaling conditions are fulfilled (see Sec.~\ref{marginal}), which will always be the case if measurements are represented as product observables, as we will discuss in more detail later in the article.

\subsection{An elastic band model}
\label{Polyelastic}

The second kind of bipartite system that we intend to consider is a macroscopic physical entity. Note that since the eighties of the last century, it was observed that the quantum laws are not the exclusive prerogative of the micro-entities, or of the very low temperature regimes, being possible to describe idealized macroscopic physical entities having a genuine quantum-like behavior, resulting from how certain non-standard experiments, perfectly well-defined in operational terms, can be performed on them, giving rise to non-Kolmogorovian probability models \citep{Aerts1986,Aertsetal1997b,Aerts1998,Aerts1999b,SassolideBianchi2013a,Sassoli2013b}. Some of these ``quantum machine'' models were also studied in order to better understand the phenomenon of entanglement, as it is possible to conceive classical laboratory situations able to violate Bell's inequalities, thus throwing some light on the possible mechanisms at the origin of the observed correlations \citep{Aerts1982,a1984,Aerts1991,aabg2000,Aerts2005,Sassoli2013b,AertsSassoli2016,aabgssv2018b}. Here we consider a model using breakable elastic bands, which is a variation of previous similar models \citep{Aerts2005,Sassoli2013b}. 

\begin{figure}[!ht]
\centering
\includegraphics[scale =0.18]{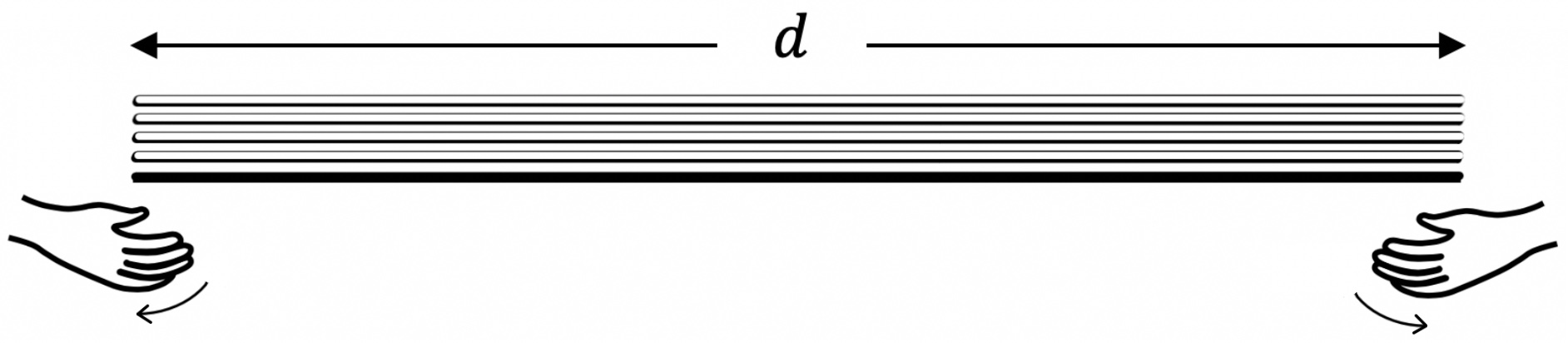}
\caption{A schematic representation of the `bundle of elastic bands' entity on which Alice and Bob perform the four $AB$, $A'B$, $AB'$ and $A'B'$ coincidence measurements.}
\label{figure2}
\end{figure}

More precisely, the entity we consider, subjected to different joint measurements, is formed by $n+1$ uniform elastic bands of same length $d$, one of which is black, and all the others are white; see Fig.~\ref{figure2}. 
It can be considered as a bipartite entity because all the elastics are aligned, parallel to each other, thus presenting all their left ends to Alice and all their right ends to Bob. The four joint measurements $AB$, $AB'$, $A'B$ and $A'B'$ are then defined as follows. Measurement $AB$ consists in Alice and Bob pulling with force, at the same predetermined time, the black elastic. Outcomes $(A_1,B_2)$ and $(A_2,B_1)$ consist in Alice collecting in this way a fragment of length greater than $d/2$ and Bob collecting a fragment of length less than $d/2$, respectively Alice collecting a fragment of length lesser than $d/2$ and Bob of length greater than $d/2$. Clearly, the elastics being assumed to be uniform, the associated probabilities are the same for Alice and Bob and equal to one-half. Also, considering that outcome $(A_1,B_1)$, where Alice and Bob both collect a fragment of length greater than $d/2$, and outcome $(A_2,B_2)$, where Alice and Bob both collect a fragment of length lesser than $d/2$, cannot be observed, we have: 
\begin{eqnarray}
&p(A_1,B_2)=p(A_2,B_1)={1\over 2},\quad p(A_1,B_1)=p(A_2,B_2)=0.
\end{eqnarray}

Measurement $AB'$ ($A'B$) is performed in the same way, but this time Bob (Alice) has to select in a random way the elastic to be pulled (for instance, keeping the eyes closed). This means that it can now either be the black one or one of the $n$ whites, and therefore Alice and Bob will not necessarily pull the same elastic. If they do not, they will simply collect the entire elastic, which therefore will be of length $d>d/2$. So, this time we have the probabilities: 
\begin{eqnarray}
&p(A'_1,B_1)=p(A_1,B'_1)={n\over n+1},\quad p(A'_2,B_2) =p(A_2,B'_2)= 0,\nonumber\\
&p(A'_1,B_2)=p(A_2,B'_1)= p(A_1,B'_2) = p(A'_2,B_1)= {1\over 2}{1\over n+1}.
\end{eqnarray}
Note that for the calculation of $p(A'_1,B_1)$, we observed that the probability for Alice not to grab the same elastic grabbed by Bob, which is the black one, is given by the number $n$ of white elastics divided by the total number $n+1$ of elastics, i.e., ${n\over n+1}$.

Finally, measurement $A'B'$ consists in Alice and Bob both pulling a random elastic, hence the probabilities are: 
\begin{eqnarray}
&p(A'_1,B'_1)={n\over n+1},\quad p(A'_2,B'_2) = 0,\quad p(A'_1,B'_2)=p(A'_2,B'_1)= {1\over 2}{1\over n+1}.
\end{eqnarray}
Note that for the calculation of $p(A'_1,B'_1)$, one has to reason as follows. There are $(n+1)^2$ ways to grab the left and right ends of $n+1$ elastics. Among these $(n+1)^2$ ways, $n+1$ of them consist in grabbing the left and right ends of the same elastic. These events have to be excluded, as they cannot produce the outcome $(A'_1,B'_1)$. So, the events where Alice and Bob grab a different elastic, which yield the $(A'_1,B'_1)$ outcome, are $(n+1)^2-(n+1)=n(n+1)$. Dividing this number by the total number $(n+1)^2$ of possible events, one obtains ${n(n+1)\over (n+1)^2}={n\over n+1}$. For the calculation of $p(A'_1,B'_2)$, one can reason as follows. Outcome $(A'_1,B'_2)$ is only possible for the $n+1$ cases where Alice and Bob grab the same elastic, and when this happens the probability is ${1\over 2}$, hence $p(A'_1,B'_2)={1\over 2}{1\over n+1}$, and the same holds of course for $p(A'_2,B'_1)$. 

We thus obtain the expectation values:
\begin{eqnarray}
&E(A,B)= -1,\quad E(A,B')=E(A',B)=E(A',B')= {n-1\over n+1}.
\label{averages-elastic}
\end{eqnarray}
Therefore, considering the quantity obtained by interchanging the roles of $B$ and $B'$ in (\ref{CHSH-quantity}), i.e., ${\rm CHSH}\equiv -E(A,B)+E(A,B')+E(A',B)+E(A',B')$, we find: 
\begin{equation}
{\rm CHSH}=1+3\, {n-1\over n+1}= 4\, {n-{1\over 2}\over n+1}.
\label{chsh-elastic}
\end{equation}
For $n=0$, we have ${\rm CHSH}=-2$, so there is no violation, as in this case all four joint measurements are the same measurement. For $n=1, 2$, we have ${\rm CHSH}=1,2$, respectively, hence there is still no violation. But for $n=3, 4, 5, 6, \dots$, we have ${\rm CHSH}={5\over 2}, {14\over 5}, 3, {22\over 7}, \dots$, respectively, so the inequality is violated for $n>2$, and will be maximally violated (${\rm CHSH}= 4$) in the $n\to \infty$ limit (in this limit, the model becomes equivalent to the `vessels of water model' introduced by one of us almost forty years ago \citep{Aerts1982}).

\subsection{A psychological model}
\label{psymodel}

Entanglement has also been identified and extensively investigated in human cognitive processes \citep{bkmm2008,bknm2009,as2011,as2014,bkrs2015,gs2017,aabgssv2018a,aabgssv2018b,Beltran2018,abgs2019}. So, as a third paradigmatic situation, we consider joint measurements performed by human participants in a psychological experiment. The system is formed by two distinct concepts \citep{as2014}: \emph{Animal} and \emph{Acts}, which are combined in a specific sentence: \emph{The Animal Acts}. This sentence carries a certain meaning, which corresponds to a specification of the state of the bipartite system formed by the two concepts \emph{Animal} and \emph{Acts}. A different choice of a sentence containing the two concepts, like for instance \emph{The Animal Acts in a Strange Way}, would correspond to a different state, producing different probabilities when performing joint measurements like those we are now going to describe. For a review of the Brussels' operational-realistic approach to cognition, where concepts are considered to be entities that can be in different states and be subjected to measurements performed by cognitive entities sensitive to their meaning, like human minds, we refer the reader to \citet{ass2016b}, and the references cited therein.

\begin{figure}[!ht]
\centering
\includegraphics[scale =0.22]{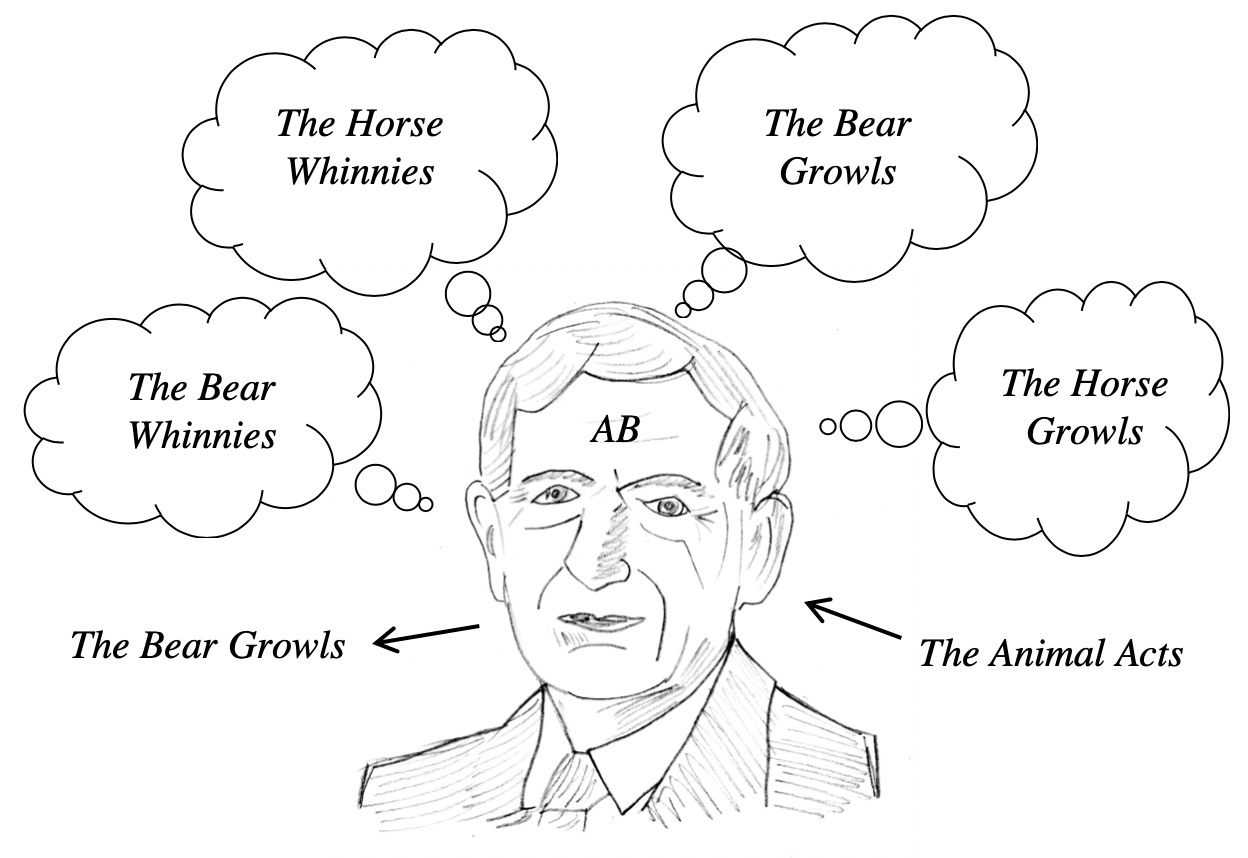}
\caption{A symbolic representation of a participant (in the appearance of David Bohm) performing the joint measurement $AB$, selecting one among the four possible combinations of exemplars, here producing a transition from the pre-measurement state \emph{The Animal Acts} to the outcome-state \emph{The Bear Growls}.}
\label{figure3}
\end{figure}

The four joint measurements $AB$, $AB'$, $A'B$ and $A'B'$ are defined as follows. Measurement $AB$ consists in participants jointly selecting a good example for the concept \emph{Animal}, from the two possibilities $A_1$ = \emph{Horse} and $A_2$ = \emph{Bear}, and a good example for the concept \emph{Acts}, from the two possibilities $B_1$ = \emph{Growls} and $B_2$ \emph{Whinnies}. Hence, the four outcomes of $AB$ are: \emph{The Horse Growls}, \emph{The Horse Whinnies}, \emph{The Bear Growls} and \emph{The Bear Whinnies}; see Fig.~\ref{figure3}.
The experimental probabilities obtained in an experiment performed in 2011 \citep{as2011} give:
\begin{eqnarray}
&p(A_1,B_1)={4\over 81},\,\, p(A_1, B_2)={51\over 81},\,\, p(A_2,B_1)={21\over 81},\,\, p(A_2,B_2)={5\over 81}.
\end{eqnarray}
The joint measurement $AB'$ is carried over in the same way, but considering for the exemplars of \emph{Acts} the two possibilities $B'_1$ = \emph{Snorts} and $B'_2$ = \emph{Meows}. Hence, the four outcomes of $AB'$ are: \emph{The Horse Snorts}, \emph{The Horse Meows}, \emph{The Bear Snorts} and \emph{The Bear Meows}, and the experimental probabilities obtained in the 2011 experiment are: 
\begin{eqnarray}
&p(A_1,B'_1)={48\over 81},\,\, p(A_1, B'_2)={2\over 81},\,\, p(A_2,B'_1)={24\over 81},\,\, p(A_2,B'_2)={7\over 81}.
\end{eqnarray}
Similarly, the joint measurement $A'B$ considers \emph{Growls} and \emph{Whinnies} for the two exemplars of \emph{Acts}, as in the $AB$ measurement, but this time $A'_1$ = \emph{Tiger} and $A'_2$ = \emph{Cat}, for the two possible choices for \emph{Animal}. Hence, the four outcomes of $A'B$ are: \emph{The Tiger Growls}, \emph{The Tiger Whinnies}, \emph{The Cat Growls} and \emph{The Cat Whinnies}, and the obtained experimental probabilities were in this case:
\begin{eqnarray}
&p(A'_1,B_1)={63\over 81},\,\, p(A'_1, B_2)={7\over 81},\,\, p(A'_2,B_1)={7\over 81},\,\, p(A'_2,B_2)={4\over 81}.
\end{eqnarray}
Finally, in joint measurement $A'B'$ the four outcomes are \emph{The Tiger Snorts}, \emph{The Tiger Meows}, \emph{The Cat Snorts} and \emph{The Cat Meows}, and the experimental probabilities obtained in the 2011 experiment are: 
\begin{eqnarray}
&p(A'_1,B'_1)={12\over 81},\,\, p(A'_1, B'_2)={7\over 81}, \,\, p(A'_2,B'_1)={8\over 81},\,\, p(A'_2,B'_2)={54\over 81}.
\end{eqnarray}
Calculating the expectation values (\ref{correlations}), using the above experimental probabilities, we obtain: 
\begin{eqnarray}
&E(A,B)= -{31\over 81},\quad E(A,B')={29\over 81}, \quad E(A',B)= {53\over 81},\quad E(A',B')= {51\over 81}.
\label{averages-psy}
\end{eqnarray}
Therefore, considering the quantity ${\rm CHSH}\equiv -E(A,B)+E(A,B')+E(A',B)+E(A',B')$, we have the violation: 
\begin{equation}
{\rm CHSH}={63\over 81} + {29\over 81} + {53\over 81} + {51\over 81} = {196\over 81}=2 + {34\over 81}>2.
\end{equation}

\section{Sub-measurements and marginal selectivity}
\label{marginal}

In the above three paradigmatic examples of bipartite systems violating the CHSH inequality, we have each time defined four joint measurements $AB$, $AB'$, $A'B$ and $A'B'$. These joint measurements are performed as a whole, in a pure coincidental way. However, one can also consider in more specific terms the sub-measurements which they are the joint of. As regards the outcome probabilities of these sub-measurements, they can be obtained by simply identifying certain outcomes of the joint measurements of which they are part. Consider the case of $AB$, whose outcomes are defined by the four couples $(A_1,B_1)$, $(A_1,B_2)$, $(A_2,B_1)$ and $(A_2,B_2)$. Let us denote by $A$ and $B$ the associated sub-measurements (we will sometimes say that these are the measurements executed by Alice and Bob, although, as we will further discuss in Sec.~\ref{more-on-submeasurements}, there are some subtleties involved when defining Alice's and Bob's sub-measurements in a fully operational way, according to the quantum formalism). 

Sub-measurement $A$ can be defined as follows: it is the measurement having the two outcomes $A_1$ and $A_2$, such that $A_1$ is actualized each time that either $(A_1,B_1)$ or $(A_1,B_2)$ are actualized, and $A_2$ is actualized each time that either $(A_2,B_1)$ or $(A_2,B_2)$ are actualized. Similarly, sub-measurement $B$ can be defined as the measurement having the two outcomes $B_1$ and $B_2$, such that $B_1$ is actualized each time that either $(A_1,B_1)$ or $(A_2,B_1)$ are actualized, and $B_2$ is actualized each time that either $(A_1,B_2)$ or $(A_2,B_2)$ are actualized. In probabilistic terms, this means that: 
\begin{eqnarray}
&p_B(A_1) = p(A_1,B_1) + p(A_1,B_2),\quad p_B(A_2) = p(A_2,B_1) + p(A_2,B_2),\nonumber\\
&p_A(B_1) = p(A_1,B_1) + p(A_2,B_1),\quad p_A(B_2) = p(A_1,B_2) + p(A_2,B_2).
\label{sub-measurementsAB}
\end{eqnarray}
Of course, we can do the same with the others three joint measurements, defining the corresponding (marginal) sub-measurement probabilities. For $AB'$ we have sub-measurement $A$ jointly performed with sub-measurement $B'$: 
\begin{eqnarray}
&p_{B'}(A_1) = p(A_1,B'_1) + p(A_1,B'_2),\quad p_{B'}(A_2) = p(A_2,B'_1) + p(A_2,B'_2),\nonumber\\
&p_A(B'_1) = p(A_1,B'_1) + p(A_2,B'_1),\quad p_A(B'_2) = p(A_1,B'_2) + p(A_2,B'_2).
\label{sub-measurementsAB'}
\end{eqnarray}
For the joint measurement $A'B$, we have sub-measurement $A'$ jointly performed with sub-measurement $B$:
\begin{eqnarray}
&p_{B}(A'_1) = p(A'_1,B_1) + p(A'_1,B_2),\quad p_{B}(A'_2) = p(A'_2,B_1) + p(A'_2,B_2),\nonumber\\
&p_{A'}(B_1) = p(A'_1,B_1) + p(A'_2,B_1),\quad p_{A'}(B_2) = p(A'_1,B_2) + p(A'_2,B_2).
\label{sub-measurementsA'B}
\end{eqnarray}
Finally, $A'B'$ is the joining of the two sub-measurement $A'$ and $B'$, with marginal outcome probabilities: 
\begin{eqnarray}
&p_{B'}(A'_1) = p(A'_1,B'_1) + p(A'_1,B'_2),\quad p_{B'}(A'_2) = p(A'_2,B'_1) + p(A'_2,B'_2),\nonumber\\
&p_{A'}(B'_1) = p(A'_1,B'_1) + p(A'_2,B'_1),\quad p_{A'}(B'_2) = p(A'_1,B'_2) + p(A'_2,B'_2).
\label{sub-measurementsA'B'}
\end{eqnarray}
The marginal laws (also called marginal selectivity, or no-signaling conditions) are then the requirements that the probabilities of the different sub-measurements do not change when one changes the sub-measurement with which they are jointly measured. More precisely, the conditions are: 
\begin{eqnarray}
&p_B(A_1)= p_{B'}(A_1),\quad p_B(A_2)= p_{B'}(A_2),\quad p_B(A'_1)= p_{B'}(A'_1),\quad p_B(A'_2)= p_{B'}(A'_2),\nonumber\\
&p_{A}(B_1) = p_{A'}(B_1),\quad p_{A}(B_2) = p_{A'}(B_2),\quad p_{A}(B'_1) = p_{A'}(B'_1),\quad p_{A}(B'_2) = p_{A'}(B'_2).
\label{marginalconditions}
\end{eqnarray}

In the example of Sec.~\ref{Bohm}, all the marginals are equal to ${1\over 2}$, hence they do obey the above marginal laws. The reason for that is that the quantum joint measurements $AB$, $AB'$, $A'B$ and $A'B'$ are associated with (tensor) product (Pauli matrix) operators $\sigma_A\otimes \sigma_B$, $\sigma_{A'}\otimes \sigma_B$, $\sigma_A\otimes \sigma_{B'}$ and $\sigma_{A'}\otimes \sigma_{B'}$, which by construction have to obey the marginal laws, as it is easy to verify by a direct calculation. Consider for instance the probability $p_B(A_1)$. Writing $\sigma_A= |A_1\rangle\langle A_1| - |A_2\rangle\langle A_2|$, and $\sigma_B= |B_1\rangle\langle B_1| - |B_2\rangle\langle B_2|$, we have: 
\begin{eqnarray}
p_B(A_1) &=& p(A_1,B_1) + p(A_1,B_2)= \langle s |\left(|A_1\rangle\langle A_1| \otimes |B_1\rangle\langle B_1|\right) |s\rangle + \langle s |\left(|A_1\rangle\langle A_1| \otimes |B_2\rangle\langle B_2|\right) |s\rangle \nonumber\\
&=& \langle s |\left(|A_1\rangle\langle A_1| \otimes (|B_1\rangle\langle B_1|+ |B_2\rangle\langle B_2|)\right) |s\rangle = \langle s |\left(|A_1\rangle\langle A_1| \otimes \mathbb{I}\right) |s\rangle,
\label{quantum-marginal}
\end{eqnarray}
which is clearly independent of $B$, hence $p_B(A_1) =p_{B'}(A_1)$, and same for the other marginal probabilities. What about the elastic example of Section~ \ref{Polyelastic}, does it obey the marginal laws? We have: 
\begin{eqnarray}
&p_B(A_1)= {1\over 2},\,\, {n+{1\over 2}\over n+1}= p_{B'}(A_1),\quad p_B(A_2)= {1\over 2},\,\, {1\over 2} {1\over n+1} = p_{B'}(A_2),\nonumber\\
&p_B(A'_1)= {n+{1\over 2}\over n+1} = p_{B'}(A'_1) ,\quad p_B(A'_2)= {1\over 2} {1\over n+1} = p_{B'}(A'_2),\nonumber\\
&p_{A}(B_1) = {1\over 2},\,\, {n+{1\over 2}\over n+1}= p_{A'}(B_1),\quad p_{A}(B_2) = {1\over 2},\,\, {1\over 2} {1\over n+1} = p_{A'}(B_2),\nonumber\\
&p_{A}(B'_1) = {n+{1\over 2}\over n+1} = p_{A'}(B'_1),\quad p_{A}(B'_2) = {1\over 2} {1\over n+1} = p_{A'}(B'_2).
\label{marginalconditions2}
\end{eqnarray}
Clearly, for the trivial case $n=0$, all marginal laws are obeyed, but for $n>0$ some of them are violated. Note that for $n=1,2$, the CHSH inequality is not violated, but the marginal laws are violated, hence the two violations are not perfectly correlated in the model. Consider however that, for example: 
\begin{eqnarray}
&p_{B'}(A_1)-p_B(A_1)={n+{1\over 2}\over n+1}-{1\over 2}=1-{1\over n+1}.
\end{eqnarray}
Hence, the degrees of the violation of the marginal laws and CHSH inequality jointly increase as $n$ increases. Let us also consider the psychological measurement of Sec.~\ref{psymodel}. We have: 
\begin{eqnarray}
&p_B(A_1)={55\over 81}\neq {50\over 81} =p_{B'}(A_1),\quad p_B(A_2)= {26\over 81}\neq {31\over 81}= p_{B'}(A_2),\nonumber\\
&p_B(A'_1)= {70\over 81}\neq {19\over 81} =p_{B'}(A'_1),\quad p_B(A'_2)= {11\over 81}\neq {62\over 81} = p_{B'}(A'_2),\nonumber\\
&p_{A}(B_1) = {25\over 81}\neq {70\over 81} = p_{A'}(B_1),\quad p_{A}(B_2) = {56\over 81}\neq {11\over 81} = p_{A'}(B_2),\nonumber\\
&p_{A}(B'_1) = {72\over 81}\neq {20\over 81} = p_{A'}(B'_1),\quad p_{A}(B'_2) = {9\over 81}\neq {61\over 81} = p_{A'}(B'_2).
\label{marginalconditions3}
\end{eqnarray}
In this example we thus have even less symmetry than in the model with the elastic bands, as all the marginal equalities are manifestly violated.

\section{Compatibility and separability}
\label{compatibility}

In the previous section we have defined the outcome probabilities of Alice's sub-measurements $A$, $A'$, and Bob's sub-measurements $B$ and $B'$, by identifying some of the outcomes of the corresponding joint measurements, as per (\ref{sub-measurementsAB})-(\ref{sub-measurementsA'B'}). These relations express the `compatibility' of the sub-measurements, in the sense that the sub-measurements behave, at least probabilistically, as if they could be substituted by one big experiment, after an identification of the corresponding outcomes. An important notion in our discussion is that of `separability'. The two compatible sub-measurements $A$ and $B$ of a joint measurement $AB$ are said to be separated, with respect to a given pre-measurement state, if the following factoring relationships holds (with the factors being independent from one another): 
\begin{eqnarray}
&p(A_1,B_1) =p(A_1)p(B_1), \quad p(A_1,B_2) =p(A_1)p(B_2), \nonumber\\
& p(A_2,B_1) =p(A_2)p(B_1),\quad p(A_2,B_2) =p(A_2)p(B_2).
\label{separability}
\end{eqnarray}
Combining (\ref{separability}) with (\ref{sub-measurementsAB}), one finds the necessary and sufficient condition for separability: 
\begin{equation}
p(A_1,B_2)p(A_2,B_1) = p(A_1,B_1)p(A_2,B_2).
\label{separability2}
\end{equation}
It is easy to check that none of the three experimental situations considered in Secs.~\ref{Bohm}-\ref{psymodel}, describe separate sub-measurements. In other words, in all these situations the joint measurements reveal correlations that violate (\ref{separability}) or (\ref{separability2}). 

Note that if all four joint measurements $AB$, $AB'$, $A'B$ and $A'B'$ are formed by separate sub-measurements, then both the CHSH inequality and the marginal laws are necessarily obeyed. This is so because then the marginal probabilities, like for instance: 
\begin{eqnarray}
p_B(A_1)&=&p(A_1,B_1)+p(A_1,B_2)=p(A_1)p(B_1)+p(A_1)p(B_2)\nonumber\\
&=&p(A_1)(p_A(B_1)+p_A(B_2))=p(A_1),
\label{marginalproof}
\end{eqnarray}
do not depend anymore on the measurement with which they are jointly executed. It is also easy to show that the CHSH is necessarily obeyed in this case. Indeed, we have:
\begin{eqnarray}
E(A,B)&=&p(A_1,B_1)-p(A_1,B_2)-p(A_2,B_1)+p(A_2,B_2)\nonumber\\
&=& p(A_1)p(B_1)-p(A_1)p(B_2)-p(A_2)p(B_1)+p(A_2)p(B_2)\\
&=& [p(A_1)-p(A_2)][p(B_1)-p(B_2)] = E(A)E(B),
\end{eqnarray}
and similarly $E(A,B')=E(A)E(B')$, $E(A',B)=E(A')E(B)$ and $E(A',B')=E(A')E(B')$. Hence: 
\begin{eqnarray}
|{\rm CHSH}| &=& |E(A,B)-E(A,B')+E(A',B)+E(A',B')|\nonumber\\
&=& |E(A)E(B)-E(A)E(B')+E(A')E(B)+E(A')E(B')|\nonumber\\
&=&|E(A)(E(B)-E(B') + E(A')(E(B)-E(B')|\nonumber\\
&\leq& |E(A)| |E(B)-E(B')| + |E(A')||E(B)-E(B')|\nonumber\\
&\leq& |E(B)-E(B')| + |E(B)-E(B')|,
\label{CHSH-proof}
\end{eqnarray}
where for the penultimate inequality we have used the triangle inequality, and for the last inequality the fact that $|E(A)|=|p(A_1)-p(A_2)| =|2p(A_1)-1|\leq 1$, and similarly $|E(A')|\leq 1$. The last expression in (\ref{CHSH-proof}) contains numbers that are within the interval $-1\leq E(B),E(B')\leq 1$, hence it is necessarily bounded by 2, i.e., $|{\rm CHSH}| \leq 2$.

\section{Correlations of the first and second kind}
\label{correlations-section}

An important aspect in the discussion of the phenomenon of entanglement (rarely taken into due account) is the distinction between `correlations of the first kind' and `correlations of the second kind' \citep{Aerts1990}. Consider the joint measurement $AB$ of Sec.~\ref{Polyelastic}, where Alice and Bob jointly pull their respective ends of the black elastic band. When they do so, the elastic (assumed to be uniform) will break with equal probability in one of its points, which neither Alice nor Bob can predict in advance. If, say, the elastic breaks at a distance $\lambda$ from Alice's end, Alice's collected fragment will be of length $L_A=\lambda$, whereas Bob's collected fragment will be of length $L_B=d-\lambda$. The remarkable property of these two lengths is that their sum $L_A+L_B$ is independent of $\lambda$ and is always equal to the initial (unstretched) length $d$ of the elastic. In other words, the two lengths $L_A$ and $L_B$ are perfectly correlated: independently of their actual value, their sum is necessarily equal to the total length $d$ of the initially unbroken elastic. So, the joint measurement $AB$ `creates a correlations' or, to say it in more precise terms, `actualizes one among an infinite number of potential correlations'. Indeed, each value of $\lambda$ corresponds to a different couple of correlated lengths $L_A$ and $L_B$, and at each run of the $AB$ measurement, because of the inevitable fluctuations in the interaction between Alice's and Bob's hands and the elastic, a different value for $\lambda$ will be obtained. Correlations that are created by a joint measurement are called `of the second kind', whereas if they were present prior to the measurement, hence are only discovered and not created by a joint measurement, they are called `of the first kind'. Consider the situation where the black elastic would be already broken, but we do not know where exactly it is broken, i.e., we do not know the exact value of $\lambda$. We only know that such ``hidden variable'' has a well-defined value prior to the joint measurement to be executed. We can for instance model our lack of knowledge by means of a uniform probability distribution, so that we still have in this case, as it was the case for the unbroken black elastic, that $p(A_1,B_2)=p(A_2,B_1)={1\over 2}$, so that $E(A,B)=-1$, and we also have $E(A',B')= {n-1\over n+1}$. What now changes are the probabilities for the $A'B$ and $AB'$ measurements. We have: 
\begin{eqnarray}
&p(A'_1,B_1)={1\over 2}{n\over n+1},\quad p(A'_1,B_2) ={1\over 2},\quad p(A'_2,B_1)={1\over 2}{1\over n+1},\quad p(A'_2,B_2)=0.
\end{eqnarray}
and similarly for the $AB'$ measurement, so that  $E(A',B) = E(A,B')=-{1\over n+1}$. For the quantity ${\rm CHSH}=-E(A,B)+E(A,B')+E(A',B)+E(A',B')$, we therefore obtain:  ${\rm CHSH}=1-{2\over n+1}+{n-1\over n+1}=2\,{n-1\over n+1}$. Clearly, for all $n$ we have $|{\rm CHSH}|\leq 2$. Considering instead ${\rm CHSH}=E(A,B)-E(A,B')+E(A',B)+E(A',B')$, we obtain: ${\rm CHSH}=-1+ {1\over n+1} -{1\over n+1} +{n-1\over n+1}=-{2\over n+1}$, so again for all $n$ we have $|{\rm CHSH}|\leq 2$. In other words, we find that when correlations of the second kind are replaced by correlations of the first kind, the CHSH inequality is not anymore violated.

The above example shows the crucial difference between an experimental situations where the lack of knowledge is about correlations (of the first kind) that pre-exist the measurement processes, which cannot give rise to a violation of the CHSH inequality, and an experimental situations where the lack of knowledge is about correlations (of the second kind) that do not pre-exist the measurements, but are created by them, during their executions. The former situation is that of so-called hidden-variable theories, introducing `elements of reality' describing those past factors that would have determined the correlations in the bipartite system, already existing prior to the measurement. These correlations of the first kind are precisely those that are filtered by Bell's inequalities, like the CHSH inequality. The latter situation is instead that of so-called hidden measurement ``interactions,'' which are assumed to be responsible for the actualization of potential properties in quantum measurements. When measurements are of the joint kind, i.e., they are performed on bipartite entities in a coincident way, at once, they can clearly also create correlations, when the bipartite entity forms an undivided whole. 

The existence of a general mechanism based on hidden measurement-interactions, responsible for quantum indeterminism, remains of course a hypothesis in need of experimental confirmation. It is important however to emphasize that such possibility is inherent in the very geometry of Hilbert space, if one adopts its `extended Bloch representation (EBR)', in which the elements of reality describing the available measurement-interactions are represented by specific simplex-structures inscribed in the Bloch sphere, allowing for a simple (non-circular) derivation of the Born rule \citep{Aerts1986,AertsSassolideBianchi2014,AertsSassoli2016}. In other words, not only the standard quantum formalism allows, but in fact also suggests, the process of actualization of potential properties, indicated by the projection postulate, to result from the presence of a hidden (non-local, non-spatial) level of interaction between the measuring system and the measured entity, which in the case of entangled systems would be responsible for the creation of correlations at the level of the whole entangled entity. 

Of course, in the elastic model, the presence of these hidden measurement-interactions is self-evident and there are no mysteries there in how the CHSH inequality and the marginal laws are violated. Different ``ways of stretching the elastic'' give rise to different non-local interactions between Alice's and Bob's hands and the elastic, which in turn will produce different breaking points, thus creating different correlations. It is important to observe that the joint action of Alice and Bob is exerted at the level of the entire elastic entity. There are no direct or indirect influences exerted by Alice's sub-measurements on Bob's sub-measurements, or vice versa, there are just global joint measurements, operated at the level of the whole bipartite structure, at once, in a non decompositional way, hence the absence of factorizability of the joint probabilities evidenced by the violation of the CHSH inequality and the additional possible violation of the marginal laws. To put it in a different way, there is nothing traveling from Alice to Bob, or from Bob to Alice, no ``spooky action at a distance,'' no superluminal communication of one part of the elastic instructing the other part how to behave. 

The situation described in the elastic model, and also subtended by the quantum formalism (when the projection postulate is taken seriously and the further elucidation provided by the EBR is considered), is also what seems to happen, mutatis mutandis, in human minds participating in a psychological experiments, like the one described in Sec.~\ref{psymodel}, where two concepts -- \emph{Animal} and \emph{Acts} -- are combined to form a bipartite entity in a given meaning state, described by the specific sentence in which the two concepts are combined, here the very simple sentence: \emph{The Animal Acts}. Even though the joint measurements are about associating different exemplars to the two conceptual entities, this operation is not performed by Alice (assuming Alice would be responsible to select the exemplar for the \emph{Animal} sub-entity) in a way that would be separated from the operation performed by Bob (assuming Bob would be responsible to select an exemplar for the \emph{Acts} sub-entity), so much so that one would be allowed to introduce notions like ``the direct or indirect influence of Alice's sub-measurement on Bob's sub-measurements, and vice versa.'' Indeed, each participant in the experiment performs the joint measurements as whole measurements, at once, considering the entire meaning of the \emph{The Animal Acts} sentence in relation to its four possible outcome-states, which in the case of the $AB$ measurement are the outcomes \emph{The Horse Growls}, \emph{The Horse Whinnies}, \emph{The Bear Growls} and \emph{The Bear Whinnies}. This is very similar to the two hands jointly pulling the two ends of the black elastic band: they are not influencing or communicating with one another, they are simply ``working in unison.'' 

Again, there is no Alice's measurement influencing (directly or indirectly) Bob's measurement, there is no action at a distance between the two sub-entities, or sub-measurements: there is just a single whole entity with a bipartite structure which is acted upon by a single measurement process -- for clarity, let us call it David's measurement -- which in a sense can be understood as the joining of Alice and Bob sub-measurements, because it implies a choice to be jointly made on \emph{Animal} and \emph{Acts}, simultaneously, in the same way that for the model of Sec.~\ref{Polyelastic}, to perform the $AB$ measurement, the black elastic needs to be jointly and simultaneously pulled from both sides. But the effective action is on the totality of the elastic entity, in the same way that David, in the psychological measurement, acts on the totality of the \emph{The Animal Acts} conceptual entity, and there would be no sense in trying to frame the experimental process in terms of possible mutual influences of Alice's action on Bob's action, and vice versa. Because, in ultimate analysis, there are no separated processes to be associated with Alice (corresponding to the choice of an exemplar for \emph{Animal}) and Bob (corresponding to the choice of an exemplar for \emph{Acts}), there is only a `single mind process', that of David, describing the collective participant in the psychological measurement, selecting at once both exemplars (fore the notion of `collective participant', see \citet{Aertsetal2019}). 

Coming back to the joint measurement $AB$ on the elastic band entity, note that from the local viewpoint of Alice and Bob, outcomes become available to them at the same time, and if the elastic is long enough, these two events will be spacelike separated. However, nothing travels at speed greater than light, as it requires time for the two elastic fragments to reach Alice's and Bob's hands (or David's hands, if the experiment is performed by a single individual). The same can be expected to be the case also with micro-physical entangled entities, like electrons and photons: even if the final detection events are spacelike separated, this does not necessarily imply that there is the propagation of a superluminal influence. Therefore, the observed violations of the marginal laws should not necessarily raise concern regarding possible violations of Einsteinian relativity. Independently of the above, the situation in the psychology laboratory is of course also different from that of a physics laboratory, as is clear that there are no principles equivalent to the relativistic one that would make one expect marginal selectivity to generally apply. This also because, when two conceptual entities are combined, and their possible exemplars are jointly selected, this only happens in the mind of a single subject, not in a spatial environment equipped with specific symmetries (see also the additional discussion in Sec.~\ref{conclusion}).

\section{Hilbertian modeling}
\label{Hilbert model}

When some of the marginal laws are violated, in addition to the CHSH inequality, we must proceed more carefully in the Hilbert space modeling of the situation. Indeed, as we showed it explicitly in the Bohm example, if the joint measurements are represented  by `product operators' in the tensor product of the Hilbert spaces of the sub-entities, the marginal laws will be satisfied (see Sec.~\ref{marginal}). However, we should not forget that in addition to introducing a tensor product starting from the state spaces of the sub-entities, it is also possible to introduce it with respect to their operator spaces, allowing in this way for a much more general tensor product construction. It can be proven that the set of linear operators on the tensor product of the state space of the joint entity is isomorphic to the tensor product of the sets of linear operators on the state spaces of the sub-entities. This means that from a mathematical perspective, when confronted with the presence of entanglement due to a violation of the CHSH inequality, there is no reason to prefer to model it in terms of an entangled state, i.e., of an element of the tensor product of the state spaces of the sub-entities, or in terms of an entangled measurement, i.e., of an element of the tensor product of the operator spaces of the sub-entities. Indeed, as we will show in the following of the article, and as we put forward in \citet{as2014,AertsSozzo2014b,AertsSozzo2014c}, when in addition to the violation of the CHSH inequality the marginal laws are also violated, this indicates that entanglement can also be present in the measurements, hence be mathematically modeled by entangled self-adjoint operators, rather than only in the state, where it can be mathematically modeled by an entangled state.

But let us analyse the situation carefully, starting from the more common way of introducing the tensor product at the level of the state spaces of the sub-systems. We can also then see very explicitly that it is indeed possible to model the situation by considering a larger class of measurements of the non-product kind, which we have called `entangled measurements' \citep{as2014,AertsSozzo2014b,AertsSozzo2014c}. In other words, when the marginal laws are violated, one cannot model entanglement only at the level of the state of the bipartite system and, as we will analyse in detail in the following, in many experimental situations some of the measurements are also to be considered as entangled. 

Let us denote by ${\cal H}$ the complex Hilbert space describing the states of the bipartite system under study. For the three examples that we considered in Sections~\ref{Bohm}-\ref{psymodel}, ${\cal H}$ is typically a 4-dimensional Hilbert space, which we will now assume to be the case, i.e., a Hilbert space whose orthonormal bases have four elements. More precisely, let us denote $x_1$, $x_2$, $x_3$ and $x_4$ the four states forming one of the basis of ${\cal H}$, which from now on we will represent as orthonormal kets $\{|x_1\rangle, \dots, |x_4\rangle\}$, $\langle x_j|x_j\rangle = \delta_{ij}$, $i,j=1,...,4$, using Dirac's notation. Let us then also consider a state $p$, represented by the normalized ket $|p\rangle\in {\cal H}$, $\langle p|p\rangle =1$. Is it an entangled state or a product state? A question of this kind cannot be answered on itself, but only with reference to a tensorial representation of ${\cal H}$. More precisely, one has first to introduce an isomorphism $I$ between ${\cal H}$ and another (isomorphic) Hilbert space ${\cal K}$, having the tensorial structure: ${\cal K}={\cal H}_A\otimes {\cal H}_B$. Then, the previous question can be more precisely stated as follows: Is the state $p$, represented by the ket $|p\rangle$, an entangled state with respect to the isomorphism $I$?

Before continuing, let us recall that an isomorphism $I$ between two Hilbert spaces ${\cal H}$ and ${\cal K}$ is a surjective linear map, $I:{\cal H}\to {\cal K}$, preserving the inner product, i.e., $\langle p|q\rangle_{\cal H}= \langle I(p) | I(q) \rangle_{\cal K}$, where $\langle \cdot|\cdot\rangle_{\cal H}$ and $\langle \cdot|\cdot \rangle_{\cal K}$ denote the scalar products in ${\cal H}$ and ${\cal K}$, respectively. Note that this automatically implies that $I$ is also injective, i.e., that $I$ is a bijection.\footnote{Note that by definition $I$ is an isometric operator and that for finite-dimensional Hilbert spaces, which will be the case in our discussion, $I$ is also necessarily a unitary operator.} So, the answer to the above question is that $p$ is an `entangled state with respect to $I$' if it is not a `product state with respect to $I$', that is, if there are no two states represented by the kets $|p_A\rangle\in {\cal H}_A$ and  $|p_B\rangle\in {\cal H}_B$, such that one could write: $I |p\rangle = |p_A\rangle\otimes |p_B\rangle$. Similarly, one can ask if a joint measurement $AB$ is an `entangled measurement', and again the question has to be addressed in relation to a specific isomorphism $I$. So, the answer is that a joint measurement $AB$, represented by a given (here bounded) self-adjoint linear operator ${\cal E}_{AB} \in {\cal L}({\cal H})$, is an `entangled measurement with respect to $I$' if it is not a `product measurement with respect to $I$', that is, if there are no two sub-measurements $A$ and $B$, represented by the self-adjoint linear operators ${\cal E}_A\in {\cal L}({\cal H}_A)$ and ${\cal E}_B\in {\cal L}({\cal H}_B)$, respectively, such that one could write: $I{\cal E}_{AB} I^{-1}= {\cal E}_A \otimes {\cal E}_B$. An important observation is that, given a joint measurement $AB$, one can always find a tailor-made isomorphism $I_{AB}: {\cal H}\to {\cal H}_A\otimes {\cal H}_B$, such that, with respect to $I_{AB}$, $AB$ is a product measurement. This is easy to prove by introducing two observables ${\cal E}_A\in {\cal L}({\cal H}_A)$ and ${\cal E}_B\in {\cal L}({\cal H}_B)$, with spectral decompositions ${\cal E}_A=a_1P_{A_1} + a_2P_{A_2}$ and ${\cal E}_B=b_1P_{B_1} + b_2P_{B_2}$, respectively, where $a_1$ and $a_2$ are the eigenvalues of ${\cal E}_A$, and $P_{A_1}=|A_1\rangle\langle A_1|$ and $P_{A_2}=|A_2\rangle\langle A_2|$ are the associated one-dimensional projection operators, and similarly for ${\cal E}_B$. We thus have:
\begin{eqnarray}
{\cal E}_A\otimes {\cal E}_B&=&(a_1P_{A_1} + a_2P_{A_2})\otimes (b_1P_{B_1} + b_2P_{B_2})\nonumber\\
&=& a_1b_1 P_{A_1}\otimes P_{B_1} + a_1b_2 P_{A_1}\otimes P_{B_2}+a_2b_1 P_{A_2}\otimes P_{B_1}+a_2b_2 P_{A_2}\otimes P_{B_2}.
\label{oaob}
\end{eqnarray}
Introducing also the spectral decomposition 
\begin{equation}
{\cal E}_{AB}=ab_{11} P_{A_1B_1} + ab_{12} P_{A_1B_2}+ab_{21} P_{A_2B_1}+ab_{22} P_{A_2B_2},
\end{equation}
where the $ab_{ij}$ and the $P_{A_iB_j}=|A_iB_j\rangle\langle A_iB_j|$, $i,j=1,2$, are the eigenvalues and associated one-dimensional projection operators of 
${\cal E}_{AB}\in {\cal L}({\cal H})$, one can then define the isomorphism $I_{AB}$ by simply specifying its action on the four eigenstates of ${\cal E}_{AB}$, or the associated projection operators:
\begin{equation}
I_{AB} |A_iB_j\rangle = |A_i\rangle\otimes |B_j\rangle, \quad I_{AB}P_{A_iB_j}I_{AB}^{-1} = P_{A_i}\otimes P_{B_j},\quad i,j=1,2.
\end{equation}
Eq.~(\ref{oaob}) can then be rewritten as: 
\begin{equation}
I_{AB}^{-1} {\cal E}_A\otimes {\cal E}_B I_{AB}= a_1b_1 P_{A_iB_1} + a_1b_2 P_{A_1B_2}+a_2b_1 P_{A_2B_1}+a_2b_2 P_{A_2B_2} = {\cal E}_{AB},
\label{oaob2}
\end{equation}
where the last equality follows from the fact that given four real numbers $ab_{ij}$, $i,j=1,2$, one can always write them as products $ab_{ij}=a_jb_j$, for well chosen $a_i$ and $b_j$, $i,j=1,2$.

Consider now two joint measurements $AB$ and $AB'$, represented by the self-adjoint operators ${\cal E}_{AB}, {\cal E}_{AB'}\in {\cal L}({\cal H})$. Let $I_{AB}$ be the isomorphism allowing to represent $AB$ as a product measurement, i.e., ${\cal E}_{AB}=I_{AB}^{-1} {\cal E}_A\otimes {\cal E}_B I_{AB}$. Can we also have ${\cal E}_{AB'}=I_{AB}^{-1}{\cal E}_A\otimes {\cal E}_{B'} I_{AB}$, i.e., can we represent both measurements $AB$ and $AB'$ as product measurements, with respect to a same isomorphism? This is what is usually considered to be the case in typical quantum situations, like in the Bohm's spin model, where a unique tensorial representation is introduced from the beginning, and all joint observables are constructed to be product observables with respect to that same tensorial representation. As we have shown already in (\ref{quantum-marginal}), when all joint measurements are product measurements, the marginal laws have to be satisfied. This means that if they are not, we cannot find a same isomorphism $I = I_{AB}=I_{AB'}$ that would allow to represent both $AB$ and $AB'$ as product measurements. Indeed: 
\begin{eqnarray}
p(A_1,B_1)+p(A_1,B_2)&=&\langle p |(P_{A_1B_1}+ P_{A_1B_2})|p \rangle = \langle p |I^{-1} I (P_{A_1B_1}+ P_{A_1B_2})I^{-1} I | p \rangle \nonumber\\
&=&\langle p |I^{-1} (P_{A_1}\otimes P_{B_1}+ P_{A_1}\otimes P_{B_2}) I | p \rangle =\langle p |I^{-1} P_{A_1}\otimes (P_{B_1}+ P_{B_2})I| p \rangle \nonumber\\
&=& \langle p |I^{-1} P_{A_1}\otimes (P_{B'_1}+ P_{B'_2})I| p \rangle = \langle p |I^{-1} (P_{A_1}\otimes P_{B'_1}+ P_{A_1}\otimes P_{B'_2}) I | p \rangle\nonumber\\
&=& \langle p |I^{-1} I (P_{A_1B'_1}+ P_{A_1B'_2})I^{-1} I | p \rangle = \langle p |(P_{A_1B'_1}+ P_{A_1B'_2})| p \rangle \nonumber\\
&=& p(A_1,B'_1)+p(A_1,B'_2),
\label{marginal-tensor}
\end{eqnarray}
i.e, the existence of such isomorphism requires the marginal laws to be satisfied. And since they are generally not, one is forced to adopt a more general representation where not all measurements are necessarily of the product form. 

In general, one can introduce four different isomorphisms, $I_{AB}$, $I_{A'B}$, $I_{AB'}$ and $I_{A'B'}$, associated with the four joint measurements $AB$, $A'B$, $AB'$ and $A'B'$, respectively, such that ${\cal E}_{AB}=I_{AB}^{-1} {\cal E}_A\otimes {\cal E}_B I_{AB}$, ${\cal E}_{A'B}=I_{A'B}^{-1} {\cal E}_{A'}\otimes {\cal E}_B I_{A'B}$, ${\cal E}_{AB'}=I_{AB'}^{-1} {\cal E}_A\otimes {\cal E}_{B'} I_{AB'}$ and ${\cal E}_{A'B'}=I_{A'B'}^{-1} {\cal E}_{A'}\otimes {\cal E}_{B'} I_{A'B'}$. In other words, $AB$ is a product measurement with respect to $I_{AB}$, $A'B$ is a product measurement with respect to $I_{A'B}$, $AB'$ is a product measurement with respect to $I_{AB'}$, and $A'B'$ is a product measurement with respect to $I_{A'B'}$. However, $AB$ will not in general be a product measurement with respect to $I_{A'B}$, $I_{AB'}$ and $I_{A'B'}$; $A'B$ will not in general be a product measurement with respect to $I_{AB}$, $I_{AB'}$ and $I_{A'B'}$; $AB'$ will not in general be a product measurement with respect to $I_{AB}$, $I_{A'B}$ and $I_{A'B'}$; and $A'B'$ will not in general be a product measurement with respect to $I_{AB}$, $I_{A'B}$ and $I_{AB'}$. To put it differently, if one wants to force a tensor product representation for all the four joint measurements $AB$, $A'B$, $AB'$ and $A'B'$, this is possible, but the price to be paid is that the representation will then become contextual, in the sense that entanglement will have to be incorporated into states that are in general different for each one of the joint measurements, i.e., the states $I_{AB}| p \rangle$, $I_{A'B}| p \rangle$, $I_{AB'}| p \rangle$ and $I_{A'B'}| p \rangle$, respectively.

\section{A simple modeling example}
\label{Elastic modeling}

It is instructive at this point to provide an example of an explicit Hilbertian modeling. For simplicity, we only consider the elastic band model presented of Sec.~\ref{Polyelastic}, in the limit $n\to\infty$, of maximal violation of the CHSH inequality. For an explicit Hilbertian representation of the psychological model of Sec.~\ref{psymodel}, we refer the reader to \citet{as2014}; see also the Hilbert space modeling provided in \citet{abgs2019}, relative to a recently performed test on the co-occurrences of two concepts and their combination in retrieval processes on specific corpuses of documents.

So, the marginal laws being violated, we cannot describe all joint measurements as product measurements with respect to a same isomorphism. Let us choose the eigenvectors of ${\cal E}_{AB}$ to be the vectors of the canonical basis of ${\cal H}={\mathbb C}^4$, that is: 
\begin{equation}
|A_1B_1\rangle = |1,0,0,0\rangle,\,\, |A_1B_2\rangle = |0,1,0,0\rangle,\,\, |A_2B_1\rangle = |0,0,1,0\rangle,\,\, |A_2B_2\rangle = |0,0,0,1\rangle. 
\end{equation}
Then, we can write the pre-measurement state as the superposition state: 
\begin{equation}
| p \rangle= {1\over \sqrt{2}}|0,1,1,0\rangle = {1\over \sqrt{2}} (|0,1,0,0\rangle+|0,0,1,0\rangle),
\end{equation}
which clearly gives the correct probabilities: 
\begin{eqnarray}
&p(A_1,B_1)=|\langle 1,0,0,0| p \rangle|^2=0,\quad 
p(A_1,B_2)=|\langle 0,1,0,0| p \rangle|^2={1\over 2},\nonumber\\
&p(A_2,B_1)=|\langle 0,0,1,0| p \rangle|^2={1\over 2},\quad 
p(A_2,B_2)=|\langle 0,0,0,1| p \rangle|^2=0.
\end{eqnarray}
Considering the isomorphism $I_{AB}$ and the observable ${\cal E}_{AB}$, defined as: 
\begin{eqnarray}
&I_{AB}|1,0,0,0\rangle = |1,0\rangle\otimes |1,0\rangle, \quad I_{AB}|0,1,0,0\rangle = |1,0\rangle\otimes |0,1\rangle,\nonumber\\
&I_{AB}|0,0,1,0\rangle = |0,1\rangle\otimes |1,0\rangle, \quad I_{AB}|0,0,0,1\rangle = |0,1\rangle\otimes |0,1\rangle,\nonumber\\
&{\cal E}_{AB}=\sum_{i,j=1}^2 ab_{ij}|A_iB_j\rangle\langle A_iB_j|,\quad ab_{ij}=\delta_i\delta_j,\quad \delta_1=1,\,\delta_2=-1,
\end{eqnarray}
such that $E(A,B)= \langle p |{\cal E}_{AB}| p \rangle$, we can write: 
\begin{eqnarray}
I_{AB}{\cal E}_{AB}I_{AB}^{-1} &=& I_{AB}(|1,0,0,0\rangle\langle 1,0,0,0|-|0,1,0,0\rangle\langle 0,1,0,0|\nonumber\\
&&\quad\quad\quad\quad -|0,0,1,0\rangle\langle 0,0,1,0|+|0,0,0,1\rangle\langle 0,0,0,1| )I_{AB}^{-1} \nonumber\\
&= &|1,0\rangle\otimes |1,0\rangle \langle 1,0|\otimes\langle 1,0| -|1,0\rangle\otimes |0,1\rangle \langle 1,0|\otimes\langle 0,1|\nonumber\\
&&\quad\quad\quad\quad - |0,1\rangle\otimes |1,0\rangle \langle 0,1|\otimes\langle 1,0| +|0,1\rangle\otimes |0,1\rangle \langle 0,1|\otimes\langle 0,1|\nonumber\\
&=&(|1,0\rangle\langle 1,0|\ - |0,1\rangle\langle 0,1|)\otimes (|1,0\rangle\langle 1,0| - |0,1\rangle\langle 0,1|) = \sigma_z\otimes\sigma_z,
\label{OAB}
\end{eqnarray}
where $\sigma_z$ is Pauli's $z$-matrix, such that $\sigma_z|1,0\rangle=|1,0\rangle$ and $\sigma_z|0,1\rangle=-|0,1\rangle$. So, the two sub-measurements $A$ and $B$ can be represented, with respect to $I_{AB}$, using the same operator ${\cal E}_A={\cal E}_B=\sigma_z$, and we have the quantum average: 
\begin{eqnarray}
E(A,B)= \langle p |{\cal E}_{AB}| p \rangle&=&\langle p |I_{AB}^{-1} I_{AB}{\cal E}_{AB}I_{AB}^{-1} I_{AB}| p \rangle = \langle p |I_{AB}^{-1} \sigma_z\otimes\sigma_z I_{AB}| p \rangle \nonumber\\
&=&{1\over 2} ( \langle 1,0|\otimes \langle 0,1| +\langle 0,1|\otimes \langle 1,0|) \sigma_z\otimes\sigma_z ( |1,0\rangle\otimes |0,1\rangle +|0,1\rangle\otimes |1,0\rangle)\nonumber\\
&=& -{1\over 2} -{1\over 2} =-1.
\label{EAB}
\end{eqnarray}

Let us now also model measurement $AB'$. We can consider in this case the eigenvectors: 
\begin{equation}
|A_1B'_1\rangle = {1\over \sqrt{2}}|0,1,1,0\rangle,\,\, |A_1B'_2\rangle ={1\over \sqrt{2}}|0,1,-1,0\rangle,\,\, |A_2B'_1\rangle = |1,0,0,0\rangle,\,\, |A_2B'_2\rangle = |0,0,0,1\rangle,
\end{equation}
which clearly give the correct probabilities: 
\begin{eqnarray}
&p(A_1,B'_1)=|{1\over \sqrt{2}}\langle 0,1,1,0| p \rangle|^2 =|\langle p | p \rangle|^2=1,\quad 
p(A_1,B'_2)=|{1\over \sqrt{2}}\langle 0,1,-1,0| p \rangle|^2=0,\nonumber\\
&p(A_2,B'_1)=|\langle 1,0,0,0| p \rangle|^2=0,\quad 
p(A_2,B'_2)=|\langle 0,0,0,1| p \rangle|^2=0.
\end{eqnarray}
For the self-adjoint operator ${\cal E}_{AB'}$, describing $AB'$, we have the spectral decomposition: 
\begin{equation}
{\cal E}_{AB'}=\sum_{i,j=1}^2 ab'_{ij}|A_iB'_j\rangle\langle A_iB'_j|,\quad ab'_{ij}=\delta_i\delta_j, \quad \delta_1=1,\,\delta_2=-1,
\end{equation}
and of course $E(A,B')= \langle p |{\cal E}_{AB'}| p \rangle=1$. However, the isomorphism $I_{AB}$ cannot be used now to also write ${\cal E}_{AB'}$ in a tensor product form, as is clear that the eigenvectors $|A_1B'_1\rangle$ and $|A_1B'_2\rangle$ are entangled vectors with respect to $I_{AB}$. 

The operators ${\cal E}_{A'B}$ and ${\cal E}_{A'B'}$ can be taken to be identical to ${\cal E}_{AB'}$, so that the Bell operator:
\begin{equation}
{\cal E}_{\rm CHSH}=-{\cal E}_{AB}+{\cal E}_{AB'}+{\cal E}_{A'B}+ {\cal E}_{A'B'}=-{\cal E}_{AB} +3{\cal E}_{AB'}
\end{equation}
can be explicitly written in the canonical basis: 
\begin{equation}
{\cal E}_{\rm CHSH} = \begin{pmatrix} 
-1 & 0 & 0 & 0 \\
0 & 1 & 0 & 0 \\
0 & 0 & 1 & 0 \\
0 & 0 & 0 & -1 
\end{pmatrix} +3 
\begin{pmatrix} 
-1 & 0 & 0 & 0 \\
0 & 0 & 1 & 0 \\
0 & 1 & 0 & 0 \\
0 & 0 & 0 & 1 
\end{pmatrix} =
\begin{pmatrix} 
-4 & 0 & 0 & 0 \\
0 & 1 & 3 & 0 \\
0 & 3 & 1 & 0 \\
0 & 0 & 0 & 2 
\end{pmatrix}.
\label{OCHSH}
\end{equation}
If we calculate the expectation value ${\rm CHSH}=\langle\ p |{\cal E}_{\rm CHSH}| p \rangle$, we thus find:
\begin{equation}
{\rm CHSH}= {1\over\sqrt{2}}
\begin{pmatrix} 
0 & 1 & 1 & 0 
\end{pmatrix} 
\begin{pmatrix} 
-4 & 0 & 0 & 0 \\
0 & 1 & 3 & 0 \\
0 & 3 & 1 & 0 \\
0 & 0 & 0 & 2 
\end{pmatrix} 
{1\over\sqrt{2}}
\begin{pmatrix} 
0 \\
1 \\
1 \\
0 
\end{pmatrix} =4,
\label{OCHSH2}
\end{equation}
in accordance with (\ref{chsh-elastic}), as $n\to\infty$.

\section{Product or entangled measurements?}
\label{product-entanglement}

In the previous sections, we have seen that entanglement, both for states and measurements, has to be defined relative to a given isomorphism, specifying a tensorial representation. This means that entanglement is a contextual property that depends on the way sub-structures are identified in a composite system. In some experimental circumstances, the way such identification has to be carried out can be dictated by the very geometry of the setting. For instance, it appears natural in a typical Bohm model situation, where Alice and Bob operate two distinct Stern-Gerlach apparatuses separated by an arbitrarily large spatial distance, to introduce a product representation for their joint measurements. However, what appears to be natural might not necessarily be correct. Indeed, in view of the numerous observed violations of the no-signaling conditions in Bell-test experiments \citep{AdenierKhrennikov2007,DeRaedt2012,DeRaedt2013,AdenierKhrennikov2016,Bednorz2017,Kupczynski2017}, one may wonder to which extent the fact that the overall measuring apparatus is formed by two spatially separated instruments, producing outcomes in a coincident way, would be sufficient to characterize all the considered joint measurements as product measurements with respect to a same tensor product representation, i.e., a same isomorphism and therefore also a same pre-measurement state. 

When entanglement was initially theorized, physicists were not expecting that correlations could manifest independently of the distance separating Alice's and Bob's measuring apparatuses. This means that when a bipartite entity is in an entangled state, even if a joint measurement is represented by means of a product operator ${\cal E}_A\otimes {\cal E}_B$, this does not mean that the corresponding sub-measurements would be separated, i.e., would obey (\ref{separability}), this being the case only when the system is in a product state. In other words, product measurements are not separated measurements, when entanglement is present in the system. Now, what we observed in Sec.~\ref{Hilbert model}, is that there is some freedom in the way such non-separability of the joint measurements (reflecting the non-separability of the bipartite entity) can be modeled within the Hilbert space formalism. At the ``local level'' of a given joint measurement, one can always contextually consider a specific isomorphism that will be able to push all the entanglement resource in the state. However, this will generally only work contextually for that specific joint measurement, and not for all joint measurements that can be defined and performed on the system. 

So, contextually, by considering a suitable isomorphism, one can always assert that `entanglement is just in the state', and not also in the joint measurement, i.e., that the notion of `entangled states' is sufficient to describe the fact that a bipartite system forms an interconnected whole, so that one does not need to also consider a notion of `entangled measurements', to properly describe the situation. As we are going to explain, this way of proceeding remains consistent only if one can consider the outcome-states to be product states, which however is something that might not be true for all experimental situations. But let us for a moment assume that the outcome-states are correctly described as product states. It becomes then possible to support the following view. First, one observes that the presence of entanglement in the system manifests at the level of the probabilities in such a way that the joint probabilities cannot be written as the products (\ref{separability}). Of course, this is a consequence of entanglement, i.e., a necessary condition for it, not a sufficient one, and it is precisely one of the merits of John Bell to have identified conditions able to characterize the presence of entanglement, via his inequalities, combining the probabilistic data obtained from different joint measurements, so as to demarcate correlations of the first kind from those of the second kind. But of course, entanglement will manifest its presence in each joint measurement, as the source of the non-factorizability of the corresponding joint probabilities. 

Now, the very notion of entanglement implicitly contains the idea that we are in the presence of a bipartite system, i.e., a system formed by two parts that, precisely, have been entangled and therefore do form a whole. Such bipartite structure can then be implemented in the formalism by describing a given joint measurement performed on the entity as a product measurement. This has the consequence that outcome-states will be modeled as product states, i.e., that the measurement process will have to be understood as a process of disentanglement of the previously entangled sub-entities (as implied by the projection postulate). So, the tensor product is to be viewed as a mathematical procedure of recognition of the bipartite structure of the system, in the context of a given joint measurement that takes the interconnected system and produces its disconnection, creating in this way the correlations. When we only focus on the situation of `one state and one joint measurement', this `tensor product procedure' presents no specific problems, as is clear that one can always find a well-defined isomorphism that can do the job of pushing all the entanglement in the state (see Sec.~\ref{Hilbert model}). However, when more than a single joint measurement is considered, and these measurements do not obey the marginal laws, we have seen that the associated isomorphisms, implementing for each of them the tensor product structure, will not coincide. When they do coincide, i.e., when the marginal laws are obeyed, one can of course forget about the different possible ways of introducing tensor product structures, and just start from the beginning with a given tensor product representation for the state space, as is usually done in quantum mechanics, where the validity of the no-signaling conditions is taken for granted. But when marginal selectivity is not satisfied, one is forced to conclude that the tensor product structure can only be locally applied to each one of the joint measurements, and not globally applied to all joint measurements. 

On could be tempted to consider the above as a shortcoming of the quantum Hilbertian formalism, however, another possibility is to simply consider this as the signature of a more complex underlying reality. Consider the example of a curved surface imbedded in ${\mathbb R}^3$. Each point of it can be locally associated with a two-dimensional tangent plane, but one cannot define a global tangent plane, i.e., a plane that would be tangent to all points of the curved surface. The situation could be here analogous: each isomorphism can locally produce a simpler tensor product structure, which however cannot be applied to all measurement situations, because the system would not possess enough symmetries for this to be possible. We could say that when the marginal laws are obeyed, the system (understood here as the entity plus its collection of relevant measurements) is as close as possible to the situation of two genuinely separated systems,\footnote{Note that because of the superposition principle, quantum mechanics does not allow for the description of fully separated systems (see \citet{Sassoli2019} and the references cited therein), which is the reason why we say ``as close as possible,'' i.e., within the limits of what the Hilbertian formalism allows to model in terms of separation.} where the presence of the sub-systems is recognizable in a much stronger way than when marginal selectivity is not satisfied. In other words, the idea is that when marginal selectivity is not satisfied, the tensor product procedure of recognition of sub-systems would not be given once for all, but needs to be fine-tuned with respect to each joint measurement considered. Otherwise, if one forces a single tensor product structure onto the system, say via the isomorphism $I_{AB}$, the other joint measurements will become, from the perspective of that $I_{AB}$, entangled measurements. However, if they produce transitions that disentangle the system (like when by pulling an elastic band it gets broken and separated into two fragments), such description by means of a single isomorphism is to be considered as non-optimal, as it makes certain measurements appear as if they would be entangled, whereas in practice they would produce a factual disentanglement of the previously entangled system. But again, their ``entangled appearance'' would only be due to the specific isomorphism used, and when reverting to a better fine-tuned isomorphism, each measurement can always be ``locally'' conveniently transformed into a product measurement, giving rise to product outcome-states (relative to that specific isomorphism). 

The above discussion is pertinent only in situations such that there is a way to ascertain that the considered joint measurements give rise to product outcome-states, i.e., states that cannot be used anymore to violate the CHSH inequality, which therefore should be properly represented as tensor product vectors. However, there certainly are situations where the outcome-states of a joint measurement are to be considered to remain entangled states.\footnote{An example of entangled measurements is the so-called Bell-state measurements, in quantum teleportation, where two given qubits are collapsed onto one of the four entangled states known as Bell states \citep{Bennet1993}.} Consider the cognitive model of Sec.~\ref{psymodel}. Take the four outcome-states of the joint measurement $AB$, described by the conceptual combinations: \emph{The Horse Growls}, \emph{The Horse Whinnies}, \emph{The Bear Growls} and \emph{The Bear Whinnies}. These are more concrete states than the pre-measurement state, \emph{The Animal Acts}, but can still be considered to be entangled states of the composite entity formed by the joining of the two individual entities \emph{Animal} and \emph{Acts}. Indeed, in the cognitive domain, what the quantum structure of entanglement captures, not only mathematically but also conceptually, is the presence of a `meaning connection' between the different concepts. The two concepts \emph{Horse} and \emph{Whinnies} are strongly meaning connected, hence one could make the case that their combination should still be represented as an entangled state, and not as a product state. The meaning connection between \emph{Bear} and \emph{Whinnies} is instead much more feeble, hence \emph{The Bear Whinnies} is to be considered to be a less entangled state than \emph{The Horse Whinnies}. Now, every time there is a meaning connection between concepts, in a given conceptual combination, one can certainly conceive joint measurements extracting correlations from it and by doing so violate the CHSH inequality. 

Therefore, for cognitive measurements the situation is that `entangled measurements' would be the default way to represent the situation of joint measurements that can truly preserve the wholeness (or part of the wholeness) of the measured conceptual entity, and not mere mathematical artefacts resulting from an inappropriate choice of the isomorphism introducing a tensor product structure.\footnote{In the domain of macroscopic models, an example of joint measurements preserving the entanglement connection was described in \citet{Sassoli2014}, with an idealized experimental situation of two prisms connected through a rigid rod, glued on their opposed polygon-faces, which were jointly rolled in order to create correlations able to violate the CHSH inequality, however preserving their interconnection through the rigid rod.} In this kind of situation, there is of course a lot of freedom in the choice of the isomorphism, as different criteria can certainly be adopted. Typically, one will consider for the pre-measurement state a maximally symmetric and entangled state, like a singlet state, so that the outcome-states of the different entangled measurements can reflect in a natural way the higher or lower meaning connection that is carried by the different combination of exemplars, which in turn translates in the fact that these outcome-states will be more or less entangled \citep{as2014, abgs2019}. So, one should not expect to find within the quantum formalism a `unique recognition procedure' in the description of the structure of a bipartite system subjected to joint measurements, i.e., a unique recipe for choosing one or multiple isomorphisms, with respect to which entanglement at the level of the states and/or measurements can be defined. Each situation requires an attentive analysis and interpretation, based on the nature of the outcome-states of the different measurements, from which a suitable representation can then be adopted.

\section{More on sub-measurements}
\label{more-on-submeasurements}

In Sec.~\ref{marginal}, we defined the two sub-measurements $A$ and $B$ of a joint measurement $AB$ by a simple procedure of identification of certain outcomes. For instance, we defined sub-measurement $A$ as the measurement having $A_i$, $i=1,2$, as its outcomes, where $A_i$ is actualized each time that either $(A_i,B_1)$ or $(A_i,B_2)$ are actualized, when performing $AB$, so that $p_B(A_i) = p(A_i,B_1) + p(A_i,B_2)$, $i=1,2$, and same for the outcomes of $B$; see (\ref{sub-measurementsAB}). In standard quantum mechanics, however, this is not how the sub-measurements of a joint measurement are usually described. Indeed, if $A$ would be just obtained by performing the ``bigger'' measurement $AB$, then simply identifying two by two its outcomes, $A$ should collapse the pre-measurement state to exactly the same set of outcome-states as $AB$ does, as operationally speaking $A$ and $AB$ would be executed in exactly the same way. 
However, using the same notation as in Sec.~\ref{Hilbert model}, the standard quantum formalism tells us to represent the two sub-measurements $A$ and $B$ by the following operators: 
\begin{eqnarray}
&&{\tilde {\cal E}}_{A}=a_{1} (P_{A_1B_1} + P_{A_1B_2})+a_{2} (P_{A_2B_1}+ P_{A_2B_2}),\nonumber\\
&&{\tilde {\cal E}}_{B}=b_{1} (P_{A_1B_1} + P_{A_2B_1})+b_{2} (P_{A_1B_2}+ P_{A_2B_2}).
\label{spectraldecompositions}
\end{eqnarray}
The ``tilded notation'' is here used to distinguish ${\tilde {\cal E}}_{A}, {\tilde {\cal E}}_{B}\in {\cal L}({\cal H})$, which act on the whole Hilbert space ${\cal H}$, from the operators ${\cal E}_A\in {\cal L}({\cal H}_A)$ and ${\cal E}_B\in {\cal L}({\cal H}_B)$, introduced in Sec.~\ref{Hilbert model}, which only act on the sub-spaces ${\cal H}_A$ and ${\cal H}_B$, in a given tensorial decomposition of ${\cal H}$. 

So, according to the projection postulate, if $| p \rangle$ is the pre-measurement state, when performing the sub-measurement $A$, represented by operator ${\tilde {\cal E}}_{A}$, the two possible outcome-states are: 
\begin{equation}
{(P_{A_1B_1} + P_{A_1B_2})| p \rangle\over \|(P_{A_1B_1} + P_{A_1B_2})| p \rangle \|},\quad {(P_{A_2B_1}+ P_{A_2B_2})| p \rangle\over \|(P_{A_2B_1}+ P_{A_2B_2})| p \rangle\|}.
\label{eigen-A}
\end{equation}
Similarly, the two possible outcome-states of sub-measurement ${\tilde {\cal E}}_{B}$ are: 
\begin{equation}
{(P_{A_1B_1} + P_{A_2B_1})| p \rangle\over \|(P_{A_1B_1} + P_{A_2B_1})| p \rangle \|},\quad {(P_{A_1B_2}+ P_{A_2B_2})| p \rangle\over \|(P_{A_1B_2}+ P_{A_2B_2})| p \rangle\|}.
\label{eigen-B}
\end{equation}
Clearly, these four outcome-states are superpositions of the four outcome-states of $AB$, which are: 
\begin{equation}
{P_{A_1B_1}| p \rangle\over \|P_{A_1B_1}| p \rangle \|},\quad {P_{A_1B_2}| p \rangle\over \|P_{A_1B_2}| p \rangle \|},\quad {P_{A_2B_1}| p \rangle\over \|P_{A_2B_1}| p \rangle \|},\quad {P_{A_2B_2}| p \rangle\over \|P_{A_2B_2}| p \rangle \|}.
\label{eigen-AB}
\end{equation}
Note however that since (see also (\ref{oaob})):
\begin{equation}
{\cal E}_{AB}={\tilde {\cal E}}_{A}{\tilde {\cal E}}_{B}={\tilde {\cal E}}_{B}{\tilde {\cal E}}_{A}= a_1b_1 P_{A_1B_1} + a_1b_2P_{A_1B_2}+a_2b_1 P_{A_2B_1}+ a_2b_2P_{A_2B_2},
\label{OAB2}
\end{equation}
even though ${\tilde {\cal E}}_{A}$ and ${\tilde {\cal E}}_{B}$ project onto different outcome-states than those of ${\cal E}_{AB}$, when considered individually, if their processes are combined the same set of outcome-states (\ref{eigen-AB}) will be consistently obtained. This can be easily seen by considering a situation where it makes sense to consider that $A$ and $B$ are performed in a sequence, one after the other (the sub-measurements being compatible, the order is irrelevant). Imagine that after performing first sub-measurement $A$, the outcome-state associated with the eigenvalue $a_1$ has been obtained (the first vector-state in (\ref{eigen-A})). If this outcome-state becomes the pre-measurement state for the subsequent measurement $B$, inserting it at the place of $| p \rangle$ in (\ref{eigen-B}), then observing that 
\begin{equation}
(P_{A_1B_1} + P_{A_2B_1})(P_{A_1B_1} + P_{A_1B_2})=P_{A_1B_1},\quad (P_{A_1B_2}+ P_{A_2B_2})(P_{A_1B_1} + P_{A_1B_2})=P_{A_1B_2},
\end{equation}
we clearly obtain the first two vectors in (\ref{eigen-AB}) as possible outcome-states. Similarly, if the outcome-state of $A$ is that associated with the eigenvalue $a_2$ (the second vector-state in (\ref{eigen-A})), observing that 
\begin{equation}
(P_{A_1B_1} + P_{A_2B_1})(P_{A_2B_1}+ P_{A_2B_2})=P_{A_2B_1},\quad (P_{A_1B_2}+ P_{A_2B_2})(P_{A_2B_1}+ P_{A_2B_2})=P_{A_2B_2},
\end{equation}
we now obtain the last two vectors in (\ref{eigen-AB}) as possible outcome-states. 

Let us now consider the example given in Sec.~\ref{psymodel}, of the psychological measurements with the bipartite conceptual entity formed by the two concepts \emph{Animal} and \emph{Acts}, in the state described by the conceptual combination \emph{The Animal Acts}. Can we specify the two sub-measurements $A$ and $B$ in more specific terms, also describing the way they individually change the initial state, so as to obtain a description in full compliance with how sub-measurements are defined in quantum mechanics? The answer is affirmative, and to show how we have to come back to the figure of David that we introduced in Sec.~\ref{correlations}, to express the fact that the joint measurements resulted from a single mind process (David being understood as the collective participant acting at once on the totality of the bipartite conceptual entity), and not from the combination of two separate mind processes, those of Alice and Bob. In other words, to see how the two sub-measurements $A$ and $B$ can be described, in a way that remains consistent with the statistics of outcomes generated by David, we have to consider some additional processes that transform the outcome-states (\ref{eigen-AB}) of $AB$ into the superposition states (\ref{eigen-A}) and (\ref{eigen-B}) corresponding to the outcome-states of $A$ and $B$, respectively. 

Starting from David's measurement $AB$, we now have to disjoin it into two sub-measurements, $A$ and $B$, associated with the figures of Alice and Bob. Clearly, the latter will have to be associated with the marginal probabilities only, and will therefore contain less information, when considered individually, than their join, hence a process of erasure must be considered to go from the whole joint measurement to the associated sub-measurements, and it is precisely this erasure (of information) process that can explain why the sub-measurements' outcome-states have to be described as superposition states (in the same way in delayed choice quantum eraser experiments one recovers the interference pattern when the ``which-path'' information is erased \citep{Yoon-Ho Kim2000,Walborn2002,Aerts2009b}). 
So, consider the situation of measurement $AB$, with David collecting the four outcomes $(A_1,B_1)$, $(A_1,B_2)$, $(A_2,B_1)$ and $(A_2,B_2)$, corresponding to the four conceptual combinations \emph{The Horse Growls}, \emph{The Horse Whinnies}, \emph{The Bear Growls} and \emph{The Bear Whinnies}. To define the sub-measurements $A$ and $B$, we have to consider two additional processes, where David sends information about the outcomes to Alice and Bob, but does so taking care to erase information about the chosen exemplar of \emph{Acts}, when informing Alice, and about the chosen exemplar for \emph{Animal}, when informing Bob. More precisely, assuming that the outcome was, say, \emph{The Horse Growls}, David will communicate the following to Alice: 
\begin{quote} 
\emph{An Exemplar For The Combination ``The Animal Acts'' Has Been Selected Among The Four Following Possibilities: ``The Horse Growls,'' ``The Horse Whinnies,'' ``The Bear Growls'' and ``The Bear Whinnies,'' And It Contained ``Horse.''}
\end{quote}
Mutatis mutandis, the same kind of communication, with removed information about what was the selection for the \emph{Animal} concept, will be given to Bob, so that Alice and Bob will be able to deduce the correct marginal probabilities, $p_B(A_1)$, $p_B(A_2)$ and $p_A(B_1)$, $p_A(B_2)$, respectively. 

Now, the above information that Alice receives from David, corresponds to an articulate conceptual combination that defines a specific state of the bipartite conceptual entity (we recall that in our Brussel's operational-realistic approach to cognition the different conceptual combinations are associated with different states for the conceptual entity under consideration; see \citet{ass2016b}). When such state is represented using a Hilbertian vector-state, it has to incorporate not only the lack of knowledge about the choice for \emph{Acts}, but also the fact that \emph{Whinnies} has greater probability to be chosen (in the human conceptual realm) than \emph{Growls}, in association with \emph{Horse}. Hence, this will correspond to a superposition state represented by the first vector in (\ref{eigen-A}), and of course one can reason in the same way for the other outcome $A_2$, and for the outcomes collected by Bob, who will receive from David a communication with the \emph{Animal} information that has been erased. 

Note that a conceptual combination expressing a situation of lack of knowledge is to be associated with a genuine new element of reality in the human conceptual realm, considering that it is human minds that constitute the measurement instruments and that the latter are sensitive to the meaning content of such combinations. So, the outcome-states associated with Alice's and Bob's sub-measurements have to be described by superposition states, and not by classical mixtures. Indeed, the additional action of David communicating the obtained outcomes to Alice and Bob, in the way described above, erasing part of the information, is equivalent to a change of state of the conceptual entity under consideration, able to produce interference effects, when subjected to additional measurements, so it truly has to be modeled as a superposition state, in accordance with the quantum projection postulate \citep{a2009a}. 

It is interesting to also note that in the above scheme there is no influence exerted by Alice on Bob, and vice versa: they both simply receive some incomplete information about the outcomes of the joint measurement creating correlations, and since the information sent to Alice is consistent with that sent to Bob, in the sense that the erased part for Alice corresponds to the non-erased part for Bob, and vice versa, and that Alice (Bob) also receive specific information on the possible outcomes of Bob's (Alice's) sub-measurements, there is no mystery as to why the marginal conditions (\ref{marginalconditions}) can be violated, when different joint measurements are considered, and of course this has no implications as regards the involvement of superluminal communication processes, even though David communicates simultaneously to Alice and Bob, sending them information by means of signals traveling, say, at the light speed (assuming that Alice and Bob would be separated by some large spatial distance).

\section{Concluding remarks}
\label{conclusion}

Let us conclude our analysis by recapitulating our findings and offering a few additional comments and contextualization. A central element and crucial insight was the recognition that in its essence the quantum entanglement phenomenon, both in physical and conceptual systems, is first of all a ``name giving'' to non-product structures that appear in a procedure of identification of what are the possible sub-entities forming a composite system. Such ``name giving'' procedure will depend on the specific choice of the isomorphism considered to implement a given bipartite tensor product structure for the space of states, and consequently also for the linear operators acting on it. A consequence of that, is that entanglement can naturally appear not only in the states, but also in the observables, i.e., in the measurements that the different observables represent. It is of course a known fact that the quantum formalism allows for the entanglement resource to be totally or partially shifted from the state to the observables, so that depending on the considered factorization a quantum state can appear either entangled or separable \citep{Thirringetal2011,HarshmanRanade2011,Harshman2012}. Our more specific point is that the interpretative freedom offered by the possibility of choosing one or multiple isomorphisms, implementing different tensor product factorizations, becomes crucial when dealing with experimental situations violating the marginal laws. According to our analysis, it is licit to affirm that in situations where the measurements' outcome-states can be assumed to remain entangled, the (non-spatial) entanglement resource should be attributed not only to the states, but also to the accessible interactions (i.e., to the measurements), operating at the level of the overall bipartite entity.

Thinking of entanglement as being present also at the level of the measurements might seem like a very drastic perspective, compared to the standard situation where it is only attributed to the state of the bipartite entity, particularly in those experimental situations where there is a clear spatial separation between the measurement apparatuses working in a coincident way. However, if the measured entity forms a whole, it is to be expected that also the measurements can become entangled, precisely through the very wholeness of the measured entity, because their action on the latter would occur simultaneously and not sequentially.\footnote{It is of course possible to also imagine experimental situations where the sub-measurements would be genuinely sequential, with mixed orders of execution. In this case, one can show that the violation of the marginal laws could result from the possible incompatibility of Alice's and Bob's experimental procedures, giving rise to order effects \citep{Sassoli2019b}.} In other words, the notions of locality and separability, usually intended as `spatial locality' and `spatial separability', need here be replaced by the more general notions of `sub-system locality' and `sub-system separability'. This because among the salient properties of physical and conceptual systems, there is precisely that of non-spatiality, and therefore `separation in space' is not anymore a sufficient criterion for characterizing a separation of two sub-systems and corresponding joint measurements. 

We have however also emphasized that it is always possible to also adopt each time, for each joint measurement, a specific tailor-made entanglement identification. Then, all the entanglement can be pushed, for each joint measurement, into the state only, with the joint measurement being described as usual as a product measurement. In this way, everything becomes explicitly contextual, in the sense that for each coincidence experiment a different state has to be used to represent the compound entity, which can only be justified when the effect of the measurement is that of disentangling the previously entangled entity. 

In Sec.~\ref{correlations-section}, we already emphasized that a violation of the marginal laws does not necessarily imply a violation of the Einsteinian no-signaling principle. It is instructive to recall here the typical reasoning that makes one believe that this is instead to be expected \citep{Ballentine1987}. One assumes that Alice and Bob have their laboratories located at great distance from one another, and that they succeeded sharing a very large number of identically prepared entangled bipartite systems, and that they are also able to jointly experiment on all of them in a parallel way, so obtaining an entire statistics of outcomes in a negligible time. Then, if all these extraordinary things can be done with great efficiency, one can imagine that Alice and Bob could have arranged things in such a way that the choice of which sub-measurement to perform, say on the part of Bob, is the expression of a code they use to communicate. And since Bob's choice of sub-measurements can be distinguished by Alice in her statistics of outcomes, because of the violation of marginal selectivity, one might conclude in this way that some kind of supraluminal communication could arise between them. 

Now, there is a loophole in the above reasoning which is usually not taken into consideration. There is no doubt that Alice has a means to infer the sub-measurement performed by Bob, if the marginal laws are violated, but what is the total duration of their overall communication? Certainly, one has to include into that duration also the time required to prepare the shared entangled entities, on which their numerous joint measurements have to be performed in parallel. These need to propagate from the source towards the two interlocutors, who we can assume to be equidistant from the latter. So, if $d$ is the distance separating Alice from Bob, in the best scenario they can collect all these shared entangled states, sent in parallel, in a time equal to ${d\over 2c}$. If we assume that their measurements are then associated with instantaneous collapses, one would be tempted to conclude that their communication can arise at an effective speed of twice the light speed. What one is here forgetting, however, is to also include the time needed for this whole process to be initiated. Indeed, the communication does not start when Alice and Bob perform their joint measurements on a statistical ensemble of entangled entities, but at the moment when they decide to activate the source in order to use it to communicate. This means that, assuming that the communication is started by Bob, he will have to send a signal to activate it, which will require at best an additional time ${d\over 2c}$. Hence, the controlled transfer of information between Alice and Bob cannot happen with effective speed greater than the speed of light. In other words, if we intend a communication as a process such that the sender of the message decides when to initiate it, and does so on purpose, then even when the marginal laws are violated one cannot exhibit a contradiction with special relativity.

In fact, additionally to the theoretical reasoning presented above, there is an even easier way to see the loophole, which is the following. Consider again the second example we put forward in this article, that of the elastics violating both the CHSH inequality and the marginal laws. We can easily make an experimental arrangement such that the events where Alice and Bob gather the outcomes of their joint measurements are spacelike separated, relativistically speaking. If the spacelike separation would be the criterion which makes it possible to use the violation of the marginal laws to send signals faster than light, according to the method sketched above, this should also work with our example of the elastics, which obviously is not the case, exactly because of the mentioned problem of the extra time needed for a complete execution of the communication. In other words, if quantum micro-entities in entangled states violating the marginal laws would lend themselves to signaling, elastic bands would be equally suitable for this purpose, which is obviously not the case. 

One might disagree with the part of the reasoning above saying that a communication between Bob and Alice can be said to have started only when one of the two has sent a signal to activate the source, but the fact remains that an effective signaling resulting from correlations of the second kind would be of a completely different nature than a causal influence of the ``spooky action at a distance'' kind, which is usually imagined to happen in situations where entanglement is at play. All sorts of correlated events happen in our physical (and cognitive) reality, which are spacelike separated, as a result of the fact that there are common causes at their origin, and nobody would of course dare to say that relativity should forbid spacelike correlated events based on common causes. Now, our proposed description of quantum measurements as processes resulting from some `hidden-measurement interactions' \citep{AertsSassolideBianchi2014,AertsSassoli2016}, is also of the kind where a common cause would be responsible for the actualization of two correlated outcomes, which can be spacelike separated, with the only (important) difference that the common cause in question would be present at the (non-spatial) potential level and would be actualized each time in a different way, at each run of a joint measurement. From that perspective, even when the marginal laws are violated, there would be no signaling in the strict sense of the notion, i.e., in the sense of having signals propagating in space with a velocity greater than that of light (see also the discussion in \citet{Aerts2014,Sassoli2019b}), so, no abandonment of the relativity principles is necessary. 

Having said that, we conclude by coming back to the criticism expressed by \citet{Dzhafarov2013}, which we already mentioned in the Introduction, according to which our analysis of the conceptual model of Sec.~\ref{psymodel} would not be revealing of the presence of a genuine form of entanglement, because the data not only violate the CHSH inequality, but also the marginal laws.\footnote{For some previous responses to these criticisms, see also \citet{Aerts2014,aabgssv2018b}.} Such criticism is based on their Contextuality-by-Default (CbD) approach, which they use to derive a modified CHSH inequality where the usual expression is corrected by subtracting terms that are non-zero if the marginal laws are violated \citep{dk2016}. This means that according to their modified inequality, a violation of the marginal laws would generally reduce the amount of entanglement (which they simply call `contextuality' in their approach) that a bipartite system is truly manifesting. This goes completely countercurrent with the view that emerges from our analysis, according to which a violation of the marginal laws, in addition to a violation of the CHSH inequality, would instead be the signature of a stronger presence of entanglement, requiring a modeling also at the level of the measurements.\footnote{This is so not only when the marginal laws are disobeyed, but also when they are obeyed but the CHSH inequality is violated beyond Tsirelson's bound \citep{as2013}.}

The reasons for this discrepancy between our interpretation and that of Dzhafarov and Kujala, is that their definition of contextuality is too restrictive to capture the overall interconnectedness that permeates both the micro-physical and cognitive realms, which give rise to the phenomenon of entanglement and which in the case of conceptual entities can be generally understood as a `connection through meaning'. Indeed, their way of looking at the experimental situation is based on the same kind of prejudice that led physicists to describe entanglement as a ``spooky action at a distance.'' For Dzhafarov and Kujala, the joint measurements described in Sec.~\ref{psymodel}, always involve two distinct interrogative contexts, one associated with Alice, who is assumed to ``select an animal,'' and the other associated with Bob, who is assumed to ``select an act,'' and these two processes are meant not to influence one another. This is of course a very unsatisfactory way of depicting the actual experimental situation, which as we explained is the result of a unique interrogative context that we described using David's fictitious character, whose cognitive process, when selecting an exemplar for \emph{The Animal Acts}, cannot be decomposed into two separate processes. Indeed, it is a process operating at the level of the non-decomposable meaning of the entire sentence, in the same way that two hands pulling an elastic band operate at the level of the entire non-decomposable unity of the elastic. And of course, this non-decompositionality of the process will generally be able to produce a violation of the CHSH inequality \emph{and} of the marginal laws, the latter being obeyed only if some remarkable symmetries would be present both in the system and measurement processes. 

In other words, Dzhafarov and Kujala's disagreement is an expression of the same prejudice that we believe is still in force today among some physicists, when they think that a `spatial separation' should also imply an `experimental separation', thus failing to recognize that our physical reality is mostly non-spatial. That a same kind of classical prejudice would be in force also among some cognitive scientists is a bit more surprising, considering that in the conceptual domain, because of the all-pervading meaning connections, the claim that an overall cognitive process should decompose into separate sub-processes is not a very natural one, for instance because a conceptual combination cannot be modelled using a joint probability distribution with its variables corresponding to the interpretation of the individual concepts. Dzhafarov and Kujala's objection is based on the observation that the choice of an exemplar for \emph{Animal} would be influenced not only by the options that are offered for that choice, but also by the options that are offered for the choice of an exemplar for \emph{Acts}. But speaking in terms of mutual influences (as is systematically done for instance in the general mathematical theory of `selective influences'; see \citet{Schweickert2012}),
means that one is already presupposing a possible separation of the two processes, i.e., a splitting of David's mind into two separate Alice's and Bob's minds, whereas in reality there is only a single-mind cognitive process or, to put it in a different way, David's mind cannot be understood as a `juxtaposition' of Alice's and Bob's sub-minds, but as their `superposition', as it also emerged from our discussion of Sec.~\ref{more-on-submeasurements}.

\end{document}